\documentclass{article}

     \PassOptionsToPackage{numbers, compress}{natbib}

\usepackage[main, final]{neurips_2026}

\usepackage[utf8]{inputenc} 
\usepackage[T1]{fontenc}    
\usepackage{hyperref}       
\usepackage{url}            
\usepackage{booktabs}       
\usepackage{amsfonts}       
\usepackage{nicefrac}       
\usepackage{microtype}      
\usepackage{xcolor}         

\title{Proximal Difference-in-Differences for Causal Learning under Confounding and Outcome Drift}

\newcommand{\Var}{\mbox{Var}}

\newcommand{\btheta}{\boldsymbol{\theta}}

\newcommand{\bbeta}{ \boldsymbol{\beta}}

\newcommand{\bpsi}{ \mbox{\boldmath $\psi$}}

\newcommand{\XX}{\checkmark}
\newcommand{\mc}{\mathcal}

\newcommand{\pre}{\textrm{pre}}

 \newcommand{\tred}{{}}
\newcommand{\ind}{\perp\!\!\!\!\perp} 
\newcommand{\E}{\mathbb{E}}
\newcommand{\pr}{\mathbb{P}}
\newcommand{\eps}{\varepsilon}

\usepackage{amsmath}
\usepackage{amssymb}
\usepackage{mathtools}
\usepackage{amsthm}
\usepackage{enumitem}

\usepackage[capitalize,noabbrev]{cleveref}

\usepackage{multirow}
\usepackage{tikz}
\usetikzlibrary{shapes,decorations,arrows,calc,arrows.meta,fit,positioning,swigs}
\tikzset{
    -Latex,auto,node distance =1 cm and 1 cm,semithick,
    state/.style ={ellipse, draw, minimum width = 0.7 cm},
    point/.style = {circle, draw, inner sep=0.04cm,fill,node contents={}},
    bidirected/.style={Latex-Latex,dashed,-},
    el/.style = {inner sep=2pt, align=left, sloped},
    swig vsplit={gap=2pt},
    swig hsplit={gap=2pt}
}

\theoremstyle{plain}
\newtheorem{theorem}{Theorem}[section]

\theoremstyle{definition}
\newtheorem{definition}[theorem]{Definition}
\newtheorem{assumption}[theorem]{Assumption}
\theoremstyle{remark}

\author{%
  Sihyung Park \\
  Department of Statistics\\
  North Carolina State University\\
  Raleigh, NC 27695 \\
  \texttt{spark52@ncsu.edu} \\
  \And
  Shu Yang\thanks{\tred{Corresponding author.} \quad $^\dagger$\,Senior author.} \\
  Department of Statistics \\
  North Carolina State University \\
  Raleigh, NC 27695 \\
  \texttt{syang24@ncsu.edu} \\
  \AND
  Ilya Lipkovich \\
  Eli Lilly and Company \\
  Indianapolis, IN 46285 \\
  \texttt{ilya.lipkovich@lilly.com} \\
  \And
  Wenyu Ye \\
  Eli Lilly and Company \\
  Indianapolis, IN 46285 \\
  \texttt{\detokenize{ye_wendy_wenyu@lilly.com}} \\
  \And
  Mingyang Shan\footnotemark[2] \\
  Eli Lilly and Company \\
  Indianapolis, IN 46285 \\
  \texttt{mingyang.shan@lilly.com} \\
}

\begin{document}

\maketitle

\begin{abstract}
  Estimating long-term treatment effects is essential in regulatory and healthcare decision-making, yet the limited follow-up duration of randomized controlled trials (RCTs) poses a significant challenge. 
Open-label extension studies attempt to capture additional information for treated subjects, but there is often significant systematic censoring and selection bias due to placebo crossover. 
While supplementing RCTs with real-world evidence is a common strategy, existing methods often rely on restrictive exchangeability assumptions that fail under latent confounding. We propose a novel proximal framework to integrate real-world evidence into an RCT with an open-label extension to estimate the long-term continued-control trajectory. Our approach successfully identifies long-term effects even in the presence of time-varying latent confounding effects and inter-study outcome drift. We provide a doubly robust locally efficient estimator that provides further resilience to model misspecification. Empirical results demonstrate that our framework maintains validity in complex data-generating processes where standard approaches may be biased.
\end{abstract}

%
\section{Introduction} \label{sec:intro}
\subsection{Background}
Estimating long-term treatment effects is critical across scientific and industrial applications. In pharmaceutical sciences, sustained efficacy and safety profiles are vital for regulatory and healthcare decision-making. 
In the technology sector, online A/B tests excel at measuring immediate user responses but often fail to capture long-term strategic goals like customer lifetime value \citep{deng2016data}. In both domains, running experiments over extended horizons is typically precluded by cost, attrition, and ethical constraints.

In clinical trials specifically, open-label extension (OLE) studies  are often added to randomized controlled trials (RCTs) to capture additional patient information after the double-blinded period of the trial. However, after the randomized double-blinded phase of an RCT, OLEs permit control group participants to cross over to the active therapy. While providing long-term longitudinal data for the treated arm, this mechanism censors the long-term counterfactual outcomes of the control arm. As a result, stakeholders must routinely extrapolate from short-term data, highlighting the necessity for robust methodological alternatives to estimate long-term counterfactuals.

To address this, emerging strategies leverage auxiliary observational data such as electronic health records in medicine or historical user logs in tech. These approaches, including propensity score integration \citep{lynch2024propensity,van2024there} and synthetic control methods \citep{abadie2010synthetic}, aim to bridge the gap between short-term experiments and long-term outcomes. However, these methods typically hinge on the strong, untestable assumption of no unmeasured confounding. The recent development of proximal causal inference offers a promising alternative, utilizing proxy variables to enable unbiased estimation even in the presence of unmeasured confounding \citep{minimalbridge, imbens2024long, miao2018identifying}.

Despite this progress, significant challenges remain. A primary limitation is the assumption of data homogeneity or no \textit{outcome drift}, which implies that counterfactual trends align between the randomized and external populations. This is rarely satisfied in practice: clinical observational cohorts often differ from patients enrolled in RCTs, while in tech, historical logs suffer from non-stationarity due to platform evolution or shifting user behavior over time. While some methods account for outcome drift, they often impose restrictive assumptions such as time-invariant confounding effects \citep{arkhangelsky2021synthetic} that are violated by common data-generating processes, or they focus on the observational population rather than the primary experimental population \citep{imbens2024long}.

\subsection{Related Work and Limitations}
For clarity of exposition, we adopt the nomenclature of clinical research, denoting the experimental cohort with extended follow-up of the treated arm as the \textit{Randomized Controlled Trial (RCT)} and the supplementary longer-horizon observational data as the \textit{External Control (EC)}. However, this formulation is directly applicable to industrial applications; for instance, in large-scale online experimentation, the ``RCT'' corresponds to a controlled A/B test, while the ``EC'' represents historical user logs leveraged to forecast long-term value.

While most approaches address some of the practical challenges, they typically assume a linear outcome-generating model or homogeneity between the RCT and EC. Even for methods that allow outcome drift, it is a common restriction to posit the absence of time-varying latent effects. Table \ref{tab:existingmethod} summarizes these requirements across the literature.

\paragraph{Synthetic Control} Synthetic control constructs a weighted combination of external control subjects to balance baseline characteristics and post-baseline outcome trajectories with the RCT control arm through the end of the double-blind period. Standard formulations \citep{abadie2021penalized, abadie2010synthetic} rely on mean exchangeability, which ignores outcome drift. While extensions \citep{arkhangelsky2021synthetic, qiu2024doubly} incorporate unit-level effects, from an asymptotic perspective, synthetic control methods require both the trial period and the EC sample size to increase indefinitely, a requirement that is restrictive in finite-sample clinical applications \citep{minimalbridge}.

\paragraph{Difference-in-Differences} Difference-in-differences (DID) relies on the parallel trends assumption to account for group-level outcome drift between RCT control and EC \citep{ghassami2022combining, zhou2024estimating}. However, the parallel trends assumption requires the influence of latent factors on the outcome to be time-invariant. This restriction precludes the modeling of dynamic, time-varying latent factors common in longitudinal clinical data.

\paragraph{Low-Rank Approximation} Matrix completion and low-rank approximation methods \citep{athey2021matrix, xiong2023large} treat long-term estimation as a data imputation problem. These approaches are typically constrained by rigid functional forms, requiring a linear relationship \citep{amjad2018robust, mao2019matrix}, which limits their utility in complex, non-linear environments.

\paragraph{Surrogates and Instrumental Variable Methods} Surrogate methods \citep{athey2019surrogate} and "bespoke" instrumental variable approaches \citep{ghassami2022combining, richardson2022bespoke} identify long-term effects by transferring short-term results through mediators. However, they generally assume no unmeasured confounding between the study enrollment and the outcome, which is frequently violated when integrating external evidence.
\tred{The method of \citet{ye2023instrumented} requires independence between the instrument and the untreated outcome trend conditional on observed covariates, the absence of an unmeasured common modifier of the instrument's effect on exposure and the exposure's effect on the outcome, and a CATE that is stable over time. These requirements limit applicability when treatment efficacy evolves over the study period.}

\paragraph{Proximal Causal Inference} Proximal causal methods use proxy variables to adjust for unmeasured confounding and identify treatment effects. Existing proximal frameworks \citep{cui2024semiparametric, minimalbridge} often require complete data and assume latent ignorability of the study cohort. While \citet{imbens2024long} relaxes latent ignorability of the study enrollment for long-term outcomes, their main identification results require three temporally ordered groups of short-term outcomes and impose external validity of the short-term potential outcomes and latent confounders.

\renewcommand{\arraystretch}{0.9}
\begin{table*}[t]
    \caption{Select methods and required assumptions. ``NUC" stands for no unmeasured confounding between the study cohort and the outcome of interest.
  ``\XX" means the assumption is strictly required; 
    ``-" means the assumption is not required.
    } \
    
    \centering
    \begin{small}
    \begin{sc}
    \begin{tabular}{c|cccccc}
        \toprule
          Methodology & NUC & Linearity & No outcome drift 
          & Target population  \\
        \midrule 
        \citet{abadie2010synthetic} & - & \XX & \XX & RCT  \\
        \citet{athey2019surrogate} & \XX & - & \XX & RCT  \\
        \citet{arkhangelsky2021synthetic} & - & \XX & - & RCT  \\ 
        \citet{athey2021matrix} & - & \XX & - & RCT  \\
        \citet{minimalbridge} & - & \XX & \XX & RCT  \\
        \citet{ghassami2022combining} & \XX & - & \XX & EC   \\
        \citet{xiong2023large} & - & \XX & \XX & RCT  \\
        \citet{zhou2024estimating} & \XX & - & - & RCT  \\
        \citet{imbens2024long} & - & - & - & EC  \\
        \bottomrule
    \end{tabular}
    \end{sc} 
    \end{small}
    \label{tab:existingmethod}
\end{table*}

\subsection{Contributions}
Our contribution centers on developing a methodology for estimating the long-term treatment effect on the experimental population via external data integration.   
This framework accommodates study-specific outcome levels and time-varying latent confounding effects and permits a proxy variable to be measured only in the EC.
Unlike a typical framework that targets the outcome directly, our approach targets the trial-to-extension trend, i.e., the change of outcome from the trial period to the extension period. This allows us to identify long-term effects even when base-level outcomes differ significantly across studies.

\begin{itemize}
    \item First, we provide a novel framework for integrating data from possibly heterogeneous RCT and EC data (Section \ref{sec:framework}). This framework accommodates latent confounding without imposing parametric models, allowing flexible machine learning methods.
    \item Second, we derive a nonparametric proximal identification strategy for the long-term counterfactual trajectory of the RCT control population (Section \ref{sec:id}). This initial identification motivates the proximal outcome regression and the proximal inverse probability weighting estimator.
    \item Third, we derive the proximal doubly robust estimator that is consistent for the counterfactual long-term outcome mean even when one of the involved nuisance functions is misspecified (Section \ref{sec:pdr}). We derive the efficient influence function of the estimand and further show that the PDR estimator is locally efficient when all nuisance models are correctly specified. \tred{We extend these results for settings when the outcome measurements are partially unobserved (Appendix \ref{AA:marl}).}
    \item Finally, the finite-sample properties and robustness of the proposed estimators are rigorously evaluated through a series of simulation studies (Section \ref{sec:numerical}). These simulations examine performance across varying degrees of model misspecification to empirically validate the theoretical consistency and efficiency gains. We further demonstrate the practical utility of the proposed estimators by applying them to a plasmode simulation, a semi-synthetic study combining real-data features with modeled outcomes, for an Alzheimer's disease trial, showcasing how this methodology can facilitate reliable long-term effect estimation in complex neurodegenerative disease research.

\end{itemize}

\section{Statistical Framework}  \label{sec:framework}
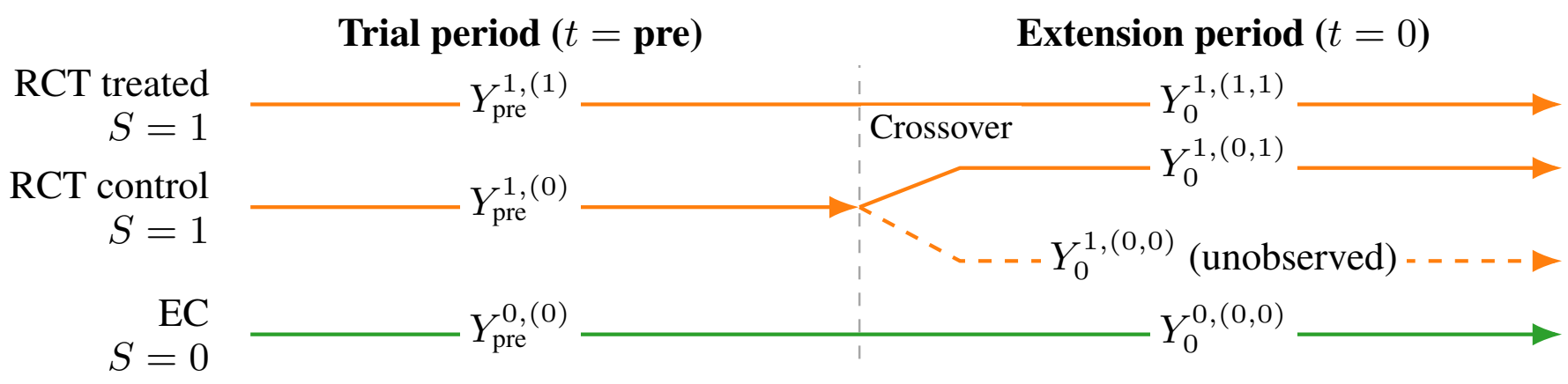
\begin{figure}[!ht]
    \centering
    \begingroup
    \setlength{\linewidth}{0.84\linewidth}
    \resizebox{\linewidth}{!}{%
    \begin{tikzpicture}[x=1cm,y=0.78cm,auto=false,
        every node/.style={font=\small,inner sep=2pt},
        observed/.style={draw=black,line width=0.85pt,-Latex},
        unobserved/.style={draw=black,dashed,line width=0.85pt,-Latex}]
        \definecolor{rctCOne}{HTML}{FF7F0E}
        \definecolor{ecCTwo}{HTML}{2CA02C}
        \tikzset{observed/.append style={draw=rctCOne},
                 unobserved/.append style={draw=rctCOne}}
        
        \path[draw=black!35,dashed,line width=0.5pt,-] (7.2,-0.65) -- (7.2,2.35);
        \node[font=\small\bfseries] at (4.5,2.65) {Trial period ($t=\pre$)};
        \node[font=\small\bfseries] at (10.1,2.65) {Extension period ($t=0$)};
        
        \node[anchor=east,align=right] at (2.1,1.95) {RCT treated\\$S=1$};
        \node[anchor=east,align=right] at (2.1,0.9) {RCT control\\$S=1$};
        \node[anchor=east,align=right] at (2.1,-0.4) {EC\\$S=0$};
        
        \path[observed] (2.35,1.95) -- (12.8,1.95);
        \node[fill=white] at (4.5,2.02) {$Y_{\pre}^{1,(1)}$};
        \node[fill=white] at (10.1,2.02) {$Y_0^{1,(1,1)}$};
        
        \path[observed] (2.35,0.9) -- (7.2,0.9);
        \node[fill=white] at (4.5,0.97) {$Y_{\pre}^{1,(0)}$};
        \path[observed] (7.2,0.9) -- (8,1.3) -- (12.8,1.3);
        \node[anchor=south,font=\footnotesize,fill=white] at (7.85,1.50) {Crossover};
        \node[fill=white] at (10.1,1.37) {$Y_0^{1,(0,1)}$};
        
        \path[unobserved] (7.2,0.9) -- (8,0.35) -- (12.8,0.35);
        \node[fill=white] at (10.1,0.42) {$Y_0^{1,(0,0)}$ (unobserved)};
        
        \begin{scope}[observed/.append style={draw=ecCTwo}]
        \path[observed] (2.35,-0.4) -- (12.8,-0.4);
        \end{scope}
        \node[fill=white] at (4.5,-0.33) {$Y_{\pre}^{0,(0)}$};
        \node[fill=white] at (10.1,-0.33) {$Y_0^{0,(0,0)}$};
    \end{tikzpicture}%
    }
    \endgroup
    \caption{\tred{RCT and EC study design: the RCT control arm crosses over to treatment during the OLE, while the EC remains untreated.}}
    \label{fig:rct_design}
\end{figure}

\subsection{Problem Statement}

We integrate an RCT with OLE ($S=1$) and EC data ($S=0$), with combined sample size $n$. Let $X$ denote baseline covariates, $U$ an unmeasured confounder, and $\bar A=(A_\pre,A_0)$ the treatment assignments during the trial ($t=\pre$) and extension ($t=0$) periods. We assume randomized assignment of $A_\pre$ in the RCT. We observe outcomes $Y_t$ for $t\in\{\pre,0\}$. The RCT provides a control arm only during the trial, while the EC extends into the extension period (Figure \ref{fig:rct_design}). This standard two-time-point framework \citep{minimalbridge,zhou2024estimating} accommodates multiple periods by aggregating trial-period and extension-period observations into composite outcomes \citep{minimalbridge,arkhangelsky2021synthetic}.

Following the convention of \citet{zhou2024estimating}, we define the counterfactual outcomes: $Y_0^{s,(a_\pre,a_0)}$ as the extension period outcome if the subject originated from $s$ and received $(a_\pre,a_0)$, and $Y_{\pre}^{s,(a_\pre)}$ as the trial period outcome if the subject originated from $s$ and received $(a_\pre)$. 
Note that $Y_{\pre}^{s,(a_\pre)} = Y_{\pre}^{s,(a_\pre, a_0)}$ for all $a_0$, since the trial period outcome is not influenced by the treatment during the extension period.
We posit the Stable Unit Treatment Value Assumption (SUTVA), asserting that the observed $Y_t$ is consistent with the assigned treatment and study cohort: $Y_\pre^{S,(A_\pre)}=Y_\pre$ and $Y_0^{S,(A_\pre,A_0)}=Y_0$.
This is a standard assumption in causal inference \citep{angrist1996identification} and implies no interference or spillover effect between units, and the implementation of the treatment and study participation remains consistent across all units.

Our primary objective is to estimate the {long-term average treatment effect (LTATE)} among the experimental population at the extension period $t=0$. This is formally defined as $
    \textrm{LTATE} = \E[Y_0^{1,(1,1)} - Y_0^{1,(0,0)} | S=1].$
We only observe specific counterfactuals from each study cohort; for subjects in the RCT, we observe ${Y}_0^{1,(1,1)}$ (treatment arm) and ${Y}_0^{1,(0,1)}$ (control arm, where crossover occurs); for subjects in the EC, we observe ${Y}_0^{0,(0,0)}$ (unexposed to the treatment).

Since the counterfactual mean $\mathbb{E}[Y_0^{1,(1,1)}|S=1] = \E[Y_0|S=1,\bar{A}=(1,1)]$ can be directly estimated from the observed data, the primary source of bias and uncertainty in estimating LTATE is the counterfactual mean $\tau := \E[Y_0^{1,(0,0)}|S=1]$. Consequently, the remainder of this article focuses on the methodology for identifying and estimating $\tau$.

\subsection{Conditional Parallel Trend Given Latent Factors}

A common assumption in data integration is {conditional exchangeability}, which posits that the untreated counterfactual mean is the same regardless of the study cohort, conditional on confounders: $\E[Y_0^{1,(0,0)}|{X},U] = \E[Y_0^{0,(0,0)}|{X},U]$ \citep{abadie2010synthetic, minimalbridge}. However, this is often violated in practice due to outcome drift. Such heterogeneity arises from differences between the data, including {placebo effects} or study bias \citep{ackerman2021generalizing, salcher2024treatment}, which result in distinct trends even for individuals with identical covariates.

The existing parallel trends assumption is commonly adopted to handle such heterogeneity. The additive equiconfounding formulation of parallel trends assumes \citep{sofer2016negative, zhou2024estimating}: for $s=0,1$,
\begin{gather}
\E[Y_{0}^{s,(0,0)} - Y_{\pre}^{s,(0)} | X,U] = \E[Y_{0}^{s,(0,0)} - Y_{\pre}^{s,(0)} | X], \label{eq:constantUeffect1} \\
\E[Y_{0}^{1,(0,0)} - Y_{\pre}^{1,(0)} | X] = \E[Y_{0}^{0,(0,0)} - Y_{\pre}^{0,(0)} | X]. \label{eq:constantUeffect2}
\end{gather}

\eqref{eq:constantUeffect1} requires the conditional mean untreated trend to be independent of $U$ after conditioning on $X$. It therefore excludes latent effects, including $X$--$U$ interactions, that change between the trial and extension periods. This limits the model's applicability, as observed covariates may interact with latent factors in ways that differentially affect outcomes across the trial and extension periods.
This constraint specifically invalidates the commonly employed linear factor model, $\mu_t({X},U) = {X}{\beta}_t + U\gamma_t$, as it necessitates that the loading $\gamma_t$ be constant over time. 
We address this limitation by introducing a relaxation, \textit{Conditional Parallel Trend given Latent factors (CPTL)}, which allows time-varying latent effects as well as $U$--$X$ interactions.

\begin{assumption}[Conditional parallel trend given latent factors]  \label{A:CPTAL}
	The conditional bias given $X$ and $U$ remains the same across time. That is,
	$
		\E[Y_t^{1,(0,0)} - Y_t^{0,(0,0)} | X,U] \equiv b(X,U)
	$ for $t \in \{0, \pre\}$.
\end{assumption}

\tred{CPTL relaxes standard assumptions by requiring trends to be parallel only after accounting for both observed and latent factors. The distinction is the level at which untreated trends are assumed transferable. Assumption \ref{A:CPTAL} permits the common conditional trend to depend on $U$ and on interactions between $X$ and $U$ through the unrestricted function $\mu_t(X,U)$ in the model above. Thus, it relaxes the combination of \eqref{eq:constantUeffect1} and \eqref{eq:constantUeffect2}. Different latent-factor distributions across cohorts can produce different trends given $X$ alone even when the trends agree given $(X,U)$, which allows the model to dispense with the restrictive condition in \eqref{eq:constantUeffect1}. }
Thus, while preserving the core benefits of the parallel trend, CPTL accommodates more realistic data-generating processes, such as:
$Y_t^{s,(0,0)} = \mu_t(X,U) + s \cdot b(X,U) + \eps_{t,s},$
where the error term $\eps_{t,s}$ is exogenous. 
Notably, if $b(X,U)=0$, this framework reduces to strict exchangeability. 
Furthermore, this structure generalizes two prominent models in panel data analysis: it simplifies to a linear factor model when the underlying functions are linear, and it reduces to an additive equiconfounding model (\eqref{eq:constantUeffect1} and \eqref{eq:constantUeffect2}) when the underlying functions are independent of $U$. 

\tred{Our framework further accommodates complex dynamics beyond existing methods.
In neurodegenerative disease research, $U$ may represent unobserved environmental or socioeconomic stressors that affect the rate of functional decline. CPTL is plausible when patients with the same observed clinical state and latent stressors have the same untreated progression rate in both cohorts, even if their initial outcome levels differ. 
For cognitive outcomes, $X$ may include education and $U$ latent intellectual capability; their interaction may affect cognitive reserve and subsequent decline. 
In digital platforms, observed demographics may modify the effect of latent seasonality or user intent on engagement trends. 
In each example, our framework permits these effects to change across phases, provided the conditional untreated trend remains common across study cohorts.}

\paragraph{Falsification test} Although the CPTL assumption is fundamentally untestable, we propose a falsification test based on a necessary condition that can be employed when multiple trial-period outcomes are available. Further details are provided in Appendix \ref{sec:cptl_test}.

\section{Nonparametric Proximal Identification} \label{sec:id}

\begin{figure}[hbt]
	\centering
	\begin{tabular}{@{}c@{\hspace{0.03\textwidth}}c@{}}
	\begin{minipage}[c]{0.45\textwidth}
		\centering
		\scalebox{.9}{
	    \begin{tikzpicture}
	    \node[state] (s) at (-2.5,0) {$S$};
	    \node[state] (x) at (0,0) {$U,X$};
	    \node[state] (w) at (2.5,0) {$W$};
	
	    \node[state] (z) at (-3.5,-1.75) {$Z$};
	    \node[state] (apre) at (-1.5,-1.75) {$A_{\pre}$};
	    \node[state] (a0) at (1.5,-1.75) {$A_0$};
	
	    \node[state] (ypre) at (-1.5,-3.5) {$Y_{\pre}$};
	    \node[state] (y) at (1.5,-3.5) {$Y_0$};
	    
	    \path (x) edge (s);
	    \path (x) edge (w);
	    \path (x) edge[bend left=3] (z);
	    \path (x) edge (apre);
	    \path (x) edge (a0);
	    \path (x) edge (ypre);
	    \path (x) edge (y);
	    
	    \path (s) edge[bidirected] (z);
	    \path (s) edge (apre);
	    \path (s) edge (a0);
	    \path (s) edge[bend right=15] (ypre);
	    \path (s) edge[bend right=0] (y);
	    
	    \path (apre) edge (ypre);
	    \path (apre) edge (a0);
	    \path (apre) edge[bidirected] (z);
	    \path (apre) edge[bend right=0] (y);
	    \path (a0) edge (y);
	    
	    \path (w) edge[bidirected, bend left=0] (ypre);
	    \path (w) edge[bidirected, bend left=15] (y);
	\end{tikzpicture}}
	\par
	\end{minipage}
	&
    \begin{minipage}[c]{0.45\textwidth}
		\centering
		\scalebox{.9}{
	    \begin{tikzpicture}
		\node[state] (x) at (0,0) {$U,X$};
		\node[state] (s) at (-1.95,0) {$S$};
	    \node[state] (w) at (1.95,0) {$W$};
	    \node[state] (ypre) at (0,-1.75) {$Y_{\pre}$};
	    \node[state] (y) at (1.35,-1.75) {$Y_0$};
	    \node[state] (z) at (-1.35,-1.75) {$Z$};
	    
	    
			
	    \path (x) edge (y);
	    \path (x) edge (s);
	    \path (x) edge (w);
	    \path (x) edge (ypre);
	    \path (x) edge[bend left=3] (z);
	    
	    \path (w) edge[bidirected, bend left=0] (ypre);
	    \path (w) edge[bidirected, bend left=0] (y);

		
		
	    \path (s) edge[bend right=0] (ypre);
	    \path (s) edge[bend right=0] (y);
	    \path (s) edge[bidirected] (z);
		\end{tikzpicture}}
	\par
	
	\end{minipage}
	\\[0.4ex]
	(a) & (b)
	\end{tabular}
	
  \caption{\tred{Causal directed acyclic graphs illustrating a possible proxy structure. (a) All variables are shown. (b) Variables are shown conditional on $\bar A=(0,0)$. Dashed edges represent associations that may or may not exist. CPTL requires equal conditional untreated counterfactual trends across studies given $X,U$.}
  }
  \label{fig:DAGs}
\end{figure}
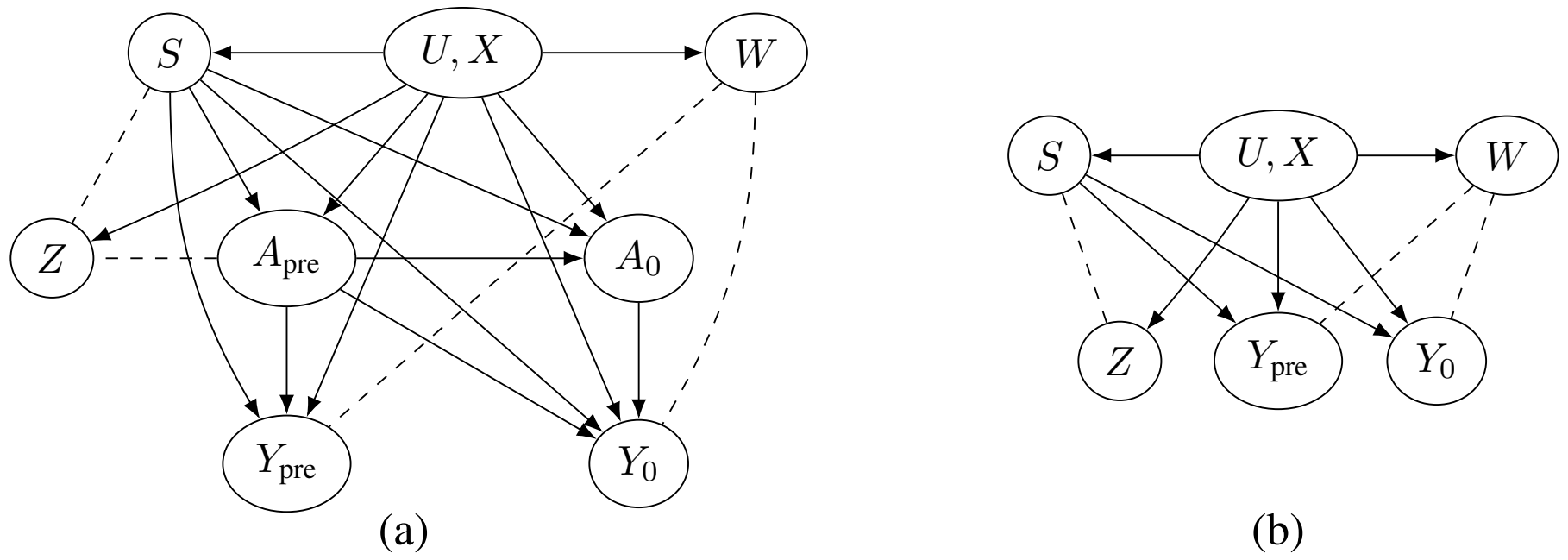

Proximal causal inference provides a technique to remove the unmeasured confounding effect on the outcome by a set of proxy variables \citep{miao2018identifying, tchetgen2024introduction}.   Suppose a {\it trend-confounding proxy} $W$ and a {\it source-confounding proxy} $Z$ exist that satisfy the following conditions.
\begin{assumption} \label{A:proxy_cond}
	$Y_0-Y_\pre \ind Z | U,X,S=0$ and $W \ind (S, Z) | U, X, A_\pre=0$.
\end{assumption} 
\tred{These conditions concern $Y_0-Y_{\pre}$ rather than each outcome level, allowing study-specific level differences under CPTL.}
Figure \ref{fig:DAGs} illustrates these conditions.

\paragraph{Guidelines for proxy selection}
In clinical research, we advise selecting $W$ from baseline comorbidities, secondary clinical measurements, or physiological biomarkers. These variables are typically recorded during the trial but are assumed to be conditionally independent of the therapy given other baseline factors. 
Conversely, we recommend choosing $Z$ from instruments or negative control variables that influence trial participation without directly affecting the outcome trend. Examples include the spatial proximity to health providers or historical referral rates at the clinical site. We also suggest that researchers consult specific trial inclusion and exclusion criteria to identify additional candidate proxies suited to their particular study context.

In the domain of online experimentation and A/B testing, we advise choosing $W$ from historical engagement metrics with unrelated platform features or baseline activity levels recorded prior to the experiment. Meanwhile, $Z$ can be chosen from exogenous technical factors such as network latency, the specific timing of exposure to a promotional banner, or engagement rates observed in historical, unrelated tests. These factors are often predictive of selection into a specific user segment or experiment but do not have a direct causal path to the long-term metric of interest.

\begin{assumption}[Latent ignorability] \label{A:ignorability} \
	$Y_t^{s,(0,0)} \ind S \mid (X,U)$ for $t=\pre,0$, $s=0,1$.
\end{assumption}

\begin{assumption}[Positivity] \label{A:positivity} \
	$\pr(S=1,\tred{A_\pre}=0 \mid X,U)>0$, $\pr(S=0\mid X,U) > 0$ almost surely over the support of $(X,U)$ in the RCT population.
\end{assumption}
\tred{Assumption \ref{A:ignorability} states that the counterfactual outcome is independent of the study membership, given all confounders. This allows confounding to be explained by both measured $X$ and unmeasured $U$, thus allowing the existence of unmeasured confounders in the analysis.
Assumption \ref{A:positivity} states that for each subject, the probability that a subject belongs to trial control or external control must be positive. This requires overlap between the studies within finer strata of $(X,U)$.}

We define a trend-confounding bridge function $d$ and a source-confounding bridge function $q$ to connect unmeasured confounding to the observed variables without explicitly modeling $U$. Here, ``source'' refers to the study cohorts. In the absence of unmeasured confounding and proxies, these simplify to the outcome regression (OR) model or inverse probability weighting (IPW).

\begin{definition}[Bridge functions] \label{def:bridge_CPTAL}
	$d(w,x)$ and $q(z,x)$ are the solutions of the integral equations:
	\begin{align}
		\E[Y_0-Y_\pre-d(W,X)|Z,X,S=0] = 0, \label{eq:d}  \\
		\E[(1-S) q(Z,X) - S\tred{(1-A_\pre)} | W,X] = 0.  \label{eq:dq}
	\end{align}
\end{definition}
Existing methods do not differentiate outcomes into transferable and non-transferable components. Our approach decouples outcomes into a non-transferable base outcome mean and a transferred mean trend. This specific decomposition, which enables heterogeneous data integration, has no analogue in prior proximal methods.
Specifically, we establish that modeling a single trend-confounding bridge function for the difference $Y_0 - Y_{\pre}$ is sufficient. A naive application of existing theory would require two separate bridge functions, one for $Y_0$ and another for $Y_{\text{pre}}$. By leveraging the trend, our approach avoids this unnecessary complexity, improving statistical efficiency and alleviating the burden of additional model specification.

Another technical but non-trivial advantage of our framework is that $Z$ does not strictly require measurement within the RCT data. Since the proxy $Z$ is consistently conditioned on the external control ($S=0$), both bridge functions can be learned with $Z$ observed only in the EC: \eqref{eq:d} conditions on $S=0$, and \eqref{eq:dq} is equivalently $\E[(1-S)q(Z,X)\mid W,X]=\pr(S=1,\tred{A_\pre=0} \mid W,X)$. 
Consequently, $W$ must be available in both cohorts, but $Z$ need not be recorded in the RCT. This reduces the trial's measurement burden, provided the proxy conditions remain substantively justified and the bridge and completeness conditions hold.

Existence of bridge functions under regularity conditions is shown in Appendix \ref{A:bridge_existence}. We discuss estimating statistical models for bridge functions in Appendix \ref{sec:learning}.

\begin{assumption}[Completeness] \label{A:completeness} \
For a square-integrable $v$, the following hold only if $v(U)=0$. \;
	(i) $\E\left[ v(U) | z,x,S=0 \right] = 0$ for all $z,x$. \;
	(ii) $\E\left[ v(U) | w,x,S=0 \right] = 0$ for all $w,x$.
\end{assumption}

Assumption \ref{A:completeness}, known as the completeness assumption, is standard in proximal identification \citep{cui2024semiparametric, miao2018identifying}.  It states that the proxies should have enough variability compared to the variability of $U$. If $U,W,Z$ are categorical, it implies that the number of categories of $W$ and $Z$ must be at least as large as that of $U$.

Given that the unmeasured confounding exerts a differential impact on $Y_t^{s,(0,0)}$ across distinct study cohorts, $d(w,x)$ connects to the untreated outcome trend, and $q(z,x)$ connects to the odds ratio of experiment participation.
The following theorem formally establishes these links between bridge functions and functionals conditional on latent factors.

\begin{theorem} \label{thm:CPTAL}
Under Assumptions \ref{A:CPTAL}, \ref{A:proxy_cond}-\ref{A:completeness}, and \ref{A:completeness_add}, $d$ and $q$ are unique solutions of
	\begin{align}
		\E[Y_0^{0,(0,0)} - Y_\pre^{0,(0)} - d(W,X)|U,X,S=0]=0, \\
		\E[ q(Z,X) | U,X,S=0 ] = \frac{\pr(S=1,\tred{A_\pre=0}|U,X)}{\pr(S=0|U,X)}.  \label{eq:qeq}
	\end{align}
\end{theorem}


%
%

While $q$ is functionally related to the odds weight in \eqref{eq:qeq}, the conditional moment equation \eqref{eq:dq} avoids a direct reciprocal formulation. This distinction provides better numerical stability during estimation by mitigating the issues typically associated with small denominators in standard inverse probability weighting approaches.

It is straightforward from the theorem that $\E[d(W,X) | S=0] = \E[Y_0^{0,(0,0)}-Y_\pre^{0,(0)} | S=0]$ and $\E[q(Z,X) | S=0] = \pr(S=1,\tred{A_\pre=0}) / \pr(S=0)$ similar to how the outcome regression and inverse probability weighting behave. The following results are motivated by this finding.

\begin{theorem}  \label{thm:pid}
Under Assumptions \ref{A:CPTAL}, \ref{A:proxy_cond}-\ref{A:completeness}, the counterfactual long-term outcome mean $\tau$ is identified:
    \begin{align*}
		{\tau}
		&= \E[ Y_\pre + d(W,X) | S=1, \tred{A_\pre=0} ]  \\
		&= \E\big[ S(1-A_\pre)Y_\pre + (1-S) q(Z,X) (Y_0-Y_\pre) \big] / \pr(S=1,A_\pre=0). 
 	\end{align*}
\end{theorem}

The first equation directly models the trend $\E[Y_0^{1,(0,0)} - Y_\pre^{1,(0)} | S=1,\tred{A_\pre=0}]$ by the mean of the trend-confounding bridge function $\E[d(W,X) | S=1,\tred{A_\pre=0}]$, and thus is referred to as the \textit{proximal outcome regression (POR)} form. The second equation, on the other hand, utilizes the source-confounding bridge function to adjust the mean EC trend for the mean RCT trend, and hence is referred to as the \textit{proximal inverse probability weighting (PIPW)} form.   We may posit models for $d(w,x), q(z,x)$ in Theorem \ref{thm:pid} and replace expectations by empirical means to obtain POR and PIPW estimators. These POR and PIPW estimators are consistent for $\tau$ by Theorem \ref{thm:pid} as long as $\hat{d}$ and $\hat{q}$ correctly estimate the true $d$ and $q$, respectively.

Under parametric specifications, a linear OR model entails a linear $d$, while a log-linear odds ratio model implies a log-linear $q$ \citep{miao2024confounding}. This correspondence allows the proximal framework to naturally leverage common parametric classes without necessitating additional complexity beyond traditional specifications.

In the absence of outcome drift, i.e., $b(X, U) = 0$, the POR and PIPW estimators recover the standard proximal estimators established in \citet{cui2024semiparametric}. Furthermore, in the absence of unmeasured confounding ($U = \emptyset$), our estimators further simplify to traditional OR and IPW methods. This methodological consistency ensures that the proposed framework maintains validity in complex longitudinal settings while remaining backward-compatible with simpler scenarios. 

\section{Proximal Doubly Robust Estimation}  \label{sec:pdr}
The consistency of POR and PIPW estimators for $\tau$ requires accurate specification of the bridge function estimates $\hat{d}$ and $\hat{q}$, respectively. This reliance on specific functional forms is often prohibitive in practice, as the underlying data-generating processes are typically unknown, rendering these estimators susceptible to bias under model misspecification.

Based on Theorem \ref{thm:pid} and semiparametric theory, we derive a doubly robust estimator and the efficient influence function (EIF) via bridge functions.   Let $\mathcal{M}_{sp}$ denote the semiparametric model that imposes no structural constraints on the observed data distribution, other than the existence of $d$.   Let $F: L_2(W,X|S=0) \to L_2(Z,X|S=0)$, $F:g \mapsto \E[g(W,X)|Z,X,S=0]$ 
be the conditional expectation operator and 
let $F': L_2(Z,X|S=0) \to L_2(W,X|S=0)$ 
defined by $F':g \mapsto {\E}[ g(Z, X) | W, X, S=0 ]$ 
be the adjoint operator of $F$.
%
We further assume the following conditions.
\begin{assumption}  \label{A:reg} \
\begin{enumerate}[topsep=0pt,itemsep=-0.5ex,partopsep=1ex,parsep=1ex]
	\item[(i)] $\hat d$ and $\hat q$ converge in probability to $d^*$ and $q^*$, respectively.
	\item[(ii)] $\pr_n\{\hat d\}$, $\pr_n\{\hat q\}$ converge in probability to $\pr\{d^*\}$, $\pr\{q^*\}$, respectively, as $n \to \infty$.
	\item[(iii)] $\hat d$ and $\hat q$ are estimated via cross-fitting, i.e., the estimated law $\hat\pr$ associated with the nuisance estimators is estimated independently of $\pr_n$.
	\item[(iv)] $\|\hat d\|$, $\|\hat q\|$ are uniformly bounded in the neighborhood $\Theta_0$ of the true nuisance functions.
\end{enumerate}
\end{assumption}
We employ cross-fitting \citep{chernozhukov2018double} to relax stringent regularity conditions. Specifically, we assume that the estimated bridge functions are uniformly $L_2$-bounded. 

Let $\E_n, \pr_n$ be the empirical mean and measure, respectively. Define the proximal doubly robust (PDR) estimator
$
	\hat{\tau}^{\textrm{PDR}} 
	= \E_n[ S\tred{(1-A_\pre)}\{ Y_\pre + \hat d(W,X) \} + (1-S) \hat q(Z,X) \hat{\delta}] 
	/ \pr_n(S=1,\tred{A_\pre=0}),  
$
where $\delta = Y_0-Y_\pre-d(W,X)$ and $\hat\delta=Y_0-Y_\pre-\hat d(W,X)$. 
%
%
The next theorem shows that the PDR estimator is consistent for $\tau$, and it additionally enjoys double robustness as well as semiparametric local efficiency.
\begin{theorem}[Proximal double machine learning] \label{thm:estimator} 		
Suppose $\pr_n(S=1,\tred{A_\pre=0}) \to \pr(S=1,\tred{A_\pre=0})$ as $n\to\infty$, and Assumptions \ref{A:CPTAL}-\ref{A:reg} and regularity conditions in the Appendix hold.
\begin{enumerate}[topsep=0pt,itemsep=-0.5ex,partopsep=1ex,parsep=1ex]
	\item[(i)] 
 	$\hat\tau^{\textrm{PDR}}$ is consistent for $\tau$ with correctly specified $\hat d$ or $\hat q$.
 	\item[(ii)] $
		\psi = [ S\tred{(1-A_\pre)}\{ Y_\pre + d(W,X) - \tau \} 
		        + (1-S)q(Z,X){\delta}]$ $/ \pr(S=1,\tred{A_\pre=0})
	$ is the EIF of $\tau$ under $\mathcal{M}_{sp}$ evaluated at the submodel that $F$ and $F'$ are surjective and \eqref{eq:dq} holds at the true data generating law.
    \item[(iii)] With correctly specified $\hat d$ and $\hat q$, $\hat\tau^{\textrm{PDR}}$ is consistent for $\tau$, and is semiparametric locally efficient, with asymptotic variance $\E[\psi^2]$, under the submodel that $F$ and $F'$ are surjective.
    \item[(iv)] Suppose $d^*=d$, $q^*=q$, and $\|F'(\hat q - q)\| \|\hat d - d\| = o_P(n^{-1/2})$ or $\|\hat q - q\| \|F(\hat d - d)\| = o_P(n^{-1/2})$ for all $\btheta \in \Theta_0$. $\hat\tau^{\textrm{PDR}}$ is consistent for $\tau$, and is semiparametric locally efficient, with asymptotic variance $\E[\psi^2]$, under the submodel that $F$ and $F'$ are surjective.
\end{enumerate}
\end{theorem}
Theorem \ref{thm:estimator} establishes double robustness and semiparametric local efficiency of the PDR estimator. Even if the primary outcome-generating process is well-understood, the corresponding bridge functions are often abstract. PDR remains consistent when either bridge function is correctly specified.
\tred{Part (iv), sometimes referred to as double machine learning \citep{chernozhukov2018double}, additionally provides a theoretical foundation for leveraging flexible, slow-converging machine learning methods to model the nuisance components in practice, particularly within the proximal learning framework.  The rate conditions in Part (iv) are closely related to the refined debiased machine learning bounds of \citet{chernozhukov2023simple}. An earlier related result for proximal causal inference is given by \citet{kallus2021causal}.}

Similar to how POR and PIPW are backward-compatible with existing methods, the PDR estimator generalizes existing strategies through a nested structure. Under $b(X,U)=0$, it recovers the estimator of \citet{cui2024semiparametric}; when additionally $U=\emptyset$, it reduces to the classical augmented IPW.

\tred{\paragraph{Extension to settings with missing outcomes} In Appendix \ref{AA:marl}, we extend this framework and these results to settings where outcome measurements are partially missing.}


%
\section{Numerical Results}  \label{sec:numerical}
\subsection{Simulation Study}  \label{sec:sim}

\begin{figure}
	\includegraphics[width=\textwidth]{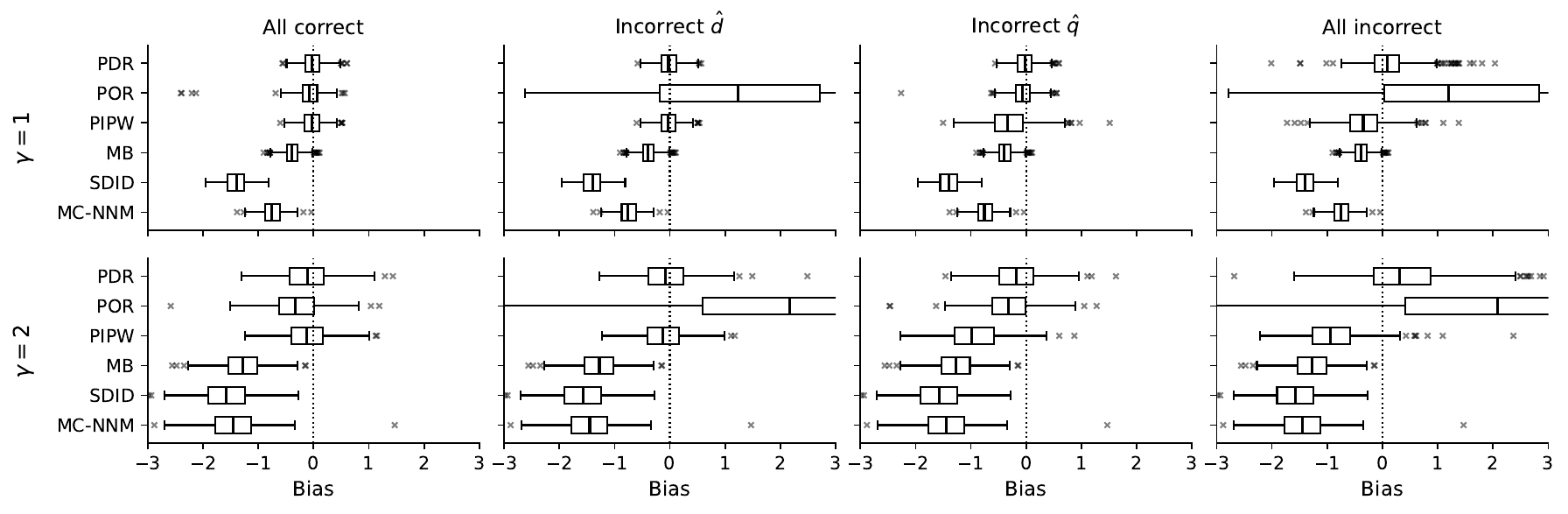}
    \centering
    \caption{\tred{Bias from simulation study for $\gamma=1$ (top) and $\gamma=2$ (bottom). The vertical dotted line indicates zero bias. For the baseline methods (MB, SDID, and MC-NNM), the simulation is unaffected by the misspecification scenarios, and their results are identical across scenarios within the same $\gamma$ value. The axes are restricted to the low-bias region for clarity.}}
    \label{fig:simres}
\end{figure}

We conducted a simulation study to evaluate the method's operating characteristics. 
The covariates $\tilde{\boldsymbol{X}}=(\tilde{X}_1, \tilde{X}_2)^\top$ were drawn from independent standard uniform distributions. The unmeasured confounder, $\tilde{U}$, was generated independently from $\mc{N}(1, 0.5)$.
We introduced non-linearity via a power transformation $\gamma$, following \citet{imbens2024long}, at values $\gamma=1$ (linear) and $\gamma=2$ (non-linear). 

$S$ was determined by a logistic model of $\boldsymbol{X} = (\tilde{X}_1^\gamma, \tilde{X}_2^\gamma)^\top$ and $U = \tilde{U}^\gamma$.
$W,Y_\pre, Y_0$, and $Z$ were generated via linear models of $\boldsymbol{X}$ and $U$, with a drift term $b(X,U)$ linear in $\boldsymbol{X}$ and $U$ added when generating $Y_\pre, Y_0$; the resulting outcome-generating process is as depicted in the model $Y_t^{s,(0,0)} = \mu_t(X,U) + s \cdot b(X,U) + \eps_{t,s}$ and Figure \ref{fig:DAGs}, representing study-specific bias. The coefficients for linear models were randomly generated, with a fixed coefficient shift between the two outcome models and a fixed random seed for the propensity model. \tred{Details of the data-generating mechanism are provided in Appendix \ref{AA:dgp}.} We generated a dataset of $N=10^6$ samples and computed the empirical mean to establish a ground-truth $\tau$.
Bridge functions were estimated using neural networks and untransformed $\tilde{\boldsymbol{X}}$ via a minimax approach  described in Appendix \ref{sec:learning}. \tred{Although sample splitting is required for the theoretical guarantees, we used the same sample for nuisance function estimation and computation of the empirical means.}

We investigated four scenarios covering all combinations of correct and incorrect specification for $\hat{d}$ and $\hat{q}$.  
We compared our framework with Synthetic Difference-in-Differences (SDID; \citep{arkhangelsky2021synthetic}), Matrix Completion via Nuclear Norm Minimization (MC-NNM; \citep{athey2021matrix}), and Minimal Bridge function estimation (MB; \citep{minimalbridge}). These methods span the current landscape of long-term treatment effect estimation on the RCT population: SDID and MC-NNM address latent confounding and outcome drift through unit fixed effects and low-rank factor structures, respectively, while MB provides a proximal identification strategy for time-varying latent effects.
The simulation was repeated $500$ times with $n=500$. Bias, root mean square error (RMSE), and coverage rate, with confidence intervals computed via stratified bootstrap, are reported.  
Figure \ref{fig:simres} and Table \ref{table:simresult1} in the Appendix summarize the findings.

The PDR estimator demonstrated the anticipated double-robustness property, provided at least one nuisance estimator, $\hat d$ or $\hat q$, was correctly specified, although modest finite-sample bias and slight undercoverage were observed when $\gamma=2$. Conversely, POR and PIPW estimators suffered from significant bias and failed to reach nominal coverage under misspecification.

In both settings, the baseline methods had larger absolute bias and lower coverage than PDR with correctly specified nuisance models. Notably, these methods fail to accurately recover the counterfactual mean in a nonlinear setting (e.g., $\gamma=2$). In contrast, our approach effectively captures these complex dependencies, maintaining robustness where traditional structural assumptions break down.


\paragraph{Sensitivity analysis} We evaluated the performance of our methodology against existing approaches under mild to moderate violations of key assumptions, as reported in Appendix \ref{sec:sensitivity}. In nearly all scenarios, our estimator maintained better performance compared to benchmarks.

\subsection{Plasmode Simulation of Alzheimer's Disease Progression}  \label{sec:plasmode}

To demonstrate the clinical utility of our framework, we construct a plasmode simulation grounded in a Phase 3 Alzheimer's Disease Trial \citep{siemers2016phase} and the GERAS observational study (\citep{wimo2013geras}; CC BY-NC-ND 4.0). The objective is to estimate the RCT control group's unobserved trend at week 52 under hypothetical patient crossover.


We standardized continuous variables and designated the baseline Mini-Mental State Examination (MMSE) score as {$U$} among available covariates, with the remaining {19 covariates} forming $\mathbf{X}$ (including BMI, age, duration of diagnosis, and cognitive and functional scores). Instrumental Activities of Daily Living (iADL) scores were selected as outcomes. \tred{Here, $Y_{\pre}$ denotes the week-28 trial-period outcome, $Y_0$ denotes the week-52 extension-period outcome, and $W$ denotes the baseline iADL score which precedes both trial and extension phases. The index $0$ denotes the extension phase, not calendar week zero.}
 
%
%
%
We generated the data to match the first six moments and correlations of the clinical data, setting the correlations between $X$ and $U$ to zero for simplicity.
We modeled the study inclusion probability with a logistic model of $\mathbf{X}$ and $U$, and $Y_t^{s,(0,0)}$ with a heterogeneous linear factor model of $\mathbf{X}$ and $U$. Following \citet{minimalbridge}, we maintained that $W$ is externally valid and thus can be used as a trend-confounding proxy, i.e., $W \ind S \mid (\mathbf{X}, U)$. The synthesized outcomes closely followed the trends in the real-world data.   We again compared our framework against MB, SDID, and MC-NNM.
Table \ref{table:lzanresult} summarizes the results, reporting bias and bootstrap standard error estimates.

\begin{table}[h]
\caption{Performance of estimators in a synthetic Alzheimer's disease study. Bias and the standard error estimate computed via a stratified bootstrap (SE) were reported.}
\label{table:lzanresult}
\centering
\begin{small}
\begin{sc}
\begin{tabular}{lrrrr}
\toprule
    & PDR    & MB   & SDID  & MC-NNM \\ 
    \midrule
    Bias  & -0.005 & -0.896 & -0.607 & -0.537 \\
    SE    &  0.478 &  0.430  & 0.974 & 0.371 \\ 
    \bottomrule
\end{tabular}
\end{sc}
\end{small}
\end{table}
Our proposed estimator yielded an estimate close to the true value while maintaining competitive precision. In contrast, the $\pm2$ SE intervals for the MB failed to cover the ground-truth trajectory. SDID and MC-NNM both underestimated the target, and SDID had the largest bootstrap SE. 

%
\section{Discussions}  \label{sec:discussion}
This work presents a framework for estimating the long-term counterfactual trajectory by integrating an RCT with an OLE and long-term observational data that may be heterogeneous. Our framework addresses latent confounding and outcome drift, and the resulting estimator also protects against model misspecification, which may be common in practice. \tred{We further extend our results to settings with missing outcomes in Appendix \ref{AA:marl}.}
To the best of our knowledge, no existing framework addresses time-varying latent effects, outcome drift, and outcome missingness simultaneously in the estimation of long-term treatment effects. 

\tred{In practice, latent confounders are rarely known. Our framework remains agnostic to their identity and ensures identification when the proxies satisfy the required conditions. 
Recent works on invalid proxies \citep{rakshit2025adaptive} and fortified proximal causal inference \citep{yu2025fortified} suggest that extending our framework to accommodate violations of the proxy assumptions could be a promising direction for future research.}

\bibliographystyle{abbrvnat}
\bibliography{example_paper}

\appendix

\section{Notation}
Throughout the appendix, we simplify the problem by assuming the only available study cohorts are external observational data ($S=0$ and $\bar A=(0,0)$) and the control arm from the experimental data during the trial period ($S=1$ and $\bar A=(0,1)$); we disregard the experimental treated arm ($S=1$ and $\bar A=(1,1)$) as it does not affect estimation and inference of the RCT control trajectory.
Let $O=\{Z,X,W,S,Y_\pre,(1-S)Y_0\}$ be the set of all observed variables.
We use $\mathcal{M}$ with subscripts ``$sp$", ``$d$", ``$q$" to denote a semiparametric model, a model with correct specification of the trend-confounding bridge function, and a model with correct specification of source-confounding bridge function, respectively. For example, under $\mathcal{M}_{d}$, we have correctly specified $d$.   
Let $\pr$ be the true distribution generating the data, $p$ be the corresponding density or mass, and $s(\cdot)$ be the score function.   Let $\pr_n$ be the empirical distribution from $\{O_i: i=1,\dots,n\}$. Denote $\pr\{\cdot\}$ by the expectation operator and $\pr_n\{\cdot\}$ by the empirical mean operator, i.e. for a functional $f$, $\pr\{f\}=\E[f(O)]$ and $\pr_n\{f\}=\E_n[f(O)]$.

\section{Existence of Bridge Functions}  \label{A:bridge_existence}
\begin{assumption} \label{A:completeness_add} For a square integrable $v$,
\begin{enumerate}
    \item[(i)] $\E[v(W) | z,x,S=0]=0$ for all $z,x$ only if $v(W)=0$.
    \item[(ii)] $\E[v(Z) | w,x,S=0]=0$ for all $w,x$ only if $v(Z)=0$.
\end{enumerate}
\end{assumption}

Let $F_{x}: L^2(p(w|x,S=0)) \to L^2(p(z|x,S=0))$, $F_{x}h = \E[h(W)|z,x,S=0]$ and $F'_{x}: L^2(p(z|x,S=0)) \to L^2(p(w|x,S=0))$, $F'_{x}q = \E[q(Z)|w,x,S=0]$.

We assume the following regularity conditions:
\begin{enumerate}
	\item[1.] $\int\int p(w|z,x,S=0) p(z|w,x,S=0) dwdz < \infty$;
	\item[2.] $\E[Y_0-Y_\pre|z,x,S=0] \in L^2(p(z|x,S=0))$;
	\item[2'.] $\frac{p(S=1|W,X)}{p(S=0|W,X)} \in L^2(p(w|x,S=0))$;
	\item[3.] $\sum_{n=1}^\infty \lambda_{x,n}^{-2}|\langle \E[Y_0-Y_\pre|z,x,S=0], \phi_{x,n} \rangle|^2 < \infty$;
	\item[3'.] $\sum_{n=1}^\infty \lambda_{x,n}^{'-2}|\langle \frac{p(S=1|W,X)}{p(S=0|W,X)}, \phi'_{x,n} \rangle|^2 < \infty$.
\end{enumerate}
By 1, $F_{x}, F'_{x}$ are compact operators for any $x$, and thus admit singular value decompositions.
Let $(\lambda_{x,n}, \varphi_{x,n}, \phi_{x,n})_{n=1}^\infty$ be a singular value decomposition of $F_{x}$.
Similarly, let $(\lambda'_{x,n}, \varphi'_{x,n}, \phi'_{x,n})_{n=1}^\infty$ be a singular value decomposition of $F'_{x}$.
The existence of solutions to \eqref{eq:d} and \eqref{eq:dq} follows immediately from Picard's theorem \citep{kress1989linear}.

\section{Training the Bridge Functions}  \label{sec:learning}
Unlike traditional OR and IPW estimators, which estimate nuisance components through direct supervised learning, bridge functions are defined as solutions to Fredholm integral equations of the first kind. This shift from pointwise regression to an inverse problem necessitates a distinct estimation strategy.

A common approach is to posit parametric models and estimate them by the generalized method of moments (GMM). This is intuitive from the conditional moment equation in Definition \ref{def:bridge_CPTAL}. By framing these integral equations as moment equations, the bridge functions can be estimated via the GMM under parametric specifications \citep{minimalbridge, miao2024confounding}.

For more flexible or high-dimensional functional classes, these functions can be learned through a minimax approach \citep{kallus2021causal}.  We obtain $\hat{d}$ and $\hat{q}$ by solving the following minimax problem:
\begin{gather*}
\hat d \in \underset{d\in\mathbb{D}}{\mathrm{argmin}} \max_{q\in\mathbb{Q}'} ( \mathbb{E}_n\left[(1-S)q(Z,X)(Y_0-Y_{\pre}-d(W,X))\right] )^2, \\
\hat q \in \underset{q\in\mathbb{Q}}{\mathrm{argmin}} \max_{d\in\mathbb{D}'} ( \mathbb{E}_n [d(W,X) \{(1-S)q(Z,X) - S\} ] )^2.
\end{gather*}
Here, $\mathbb{D}$ and $\mathbb{Q}$ denote the classes for the bridge functions, while $\mathbb{D}'$ and $\mathbb{Q}'$ represent the spaces of ``critic'' functions. In this adversarial framework, the critics are trained to identify the most severe violations of the identifying moment conditions, while the bridge functions are optimized to minimize these worst-case residuals.

\citet{kallus2021causal} showed that when the critic classes are defined as closed balls within Reproducing Kernel Hilbert Spaces induced by positive-semidefinite kernels $k_w$ and $k_z$, the minimax problem yields a closed-form loss. Specifically, the bridge functions are estimated by minimizing the kernel-induced norm of the conditional moment restrictions:
\begin{align}
\hat d &\in \underset{d\in\mathbb{D}}{\mathrm{argmin}} \big[ \{{\bf y}_0 \ominus {\bf y}_\pre \ominus d({\bf w},{\bf x})\}^\top {\bf K}_{z,n} \{{\bf y}_0 \ominus {\bf y}_\pre \ominus d({\bf w},{\bf x})\} \big],  \label{eq:minimax_d} \\
\hat q &\in \underset{q\in\mathbb{Q}}{\mathrm{argmin}} \big[ \{ ({\bf1} \ominus {\bf s}) \odot q({\bf z},{\bf x}) \ominus {\bf s} \}^\top {\bf K}_{w,n} \{ ({\bf1} \ominus {\bf s}) \odot q({\bf z},{\bf x}) \ominus {\bf s} \}  \big],  \label{eq:minimax_q}
\end{align}
where ${\bf y}_0, {\bf y}_\pre, \text{ and } {\bf s}$ are the observed outcome and study indicator vectors and ${\bf w}, {\bf z}, {\bf x}$ are the proxy and covariate vectors, while $\ominus$ and $\odot$ denote element-wise subtraction and multiplication, ${\bf K}_{z,n}$ is a matrix whose $(i,j)$th entry is $(1-S_i)(1-S_j)k_z((Z_i,X_i), (Z_j,X_j))$, and ${\bf K}_{w,n}$ is a matrix whose $(i,j)$th entry is $k_w((W_i,X_i), (W_j,X_j))$. For a general, rigorous discussion of this approach, we refer the reader to \citet{kallus2021causal}.

For the numerical study in Section \ref{sec:sim}, we employ a linear kernel for $k_w$ and a radial basis function kernel for $k_z$. We parameterized the bridge functions using feed-forward neural networks. For $\hat{d}$, we employed an architecture with two hidden layers ($20$ and $10$ units), utilizing ReLU activations and a linear output layer. $\hat{q}$ is modeled with another ($20, 10$) hidden layer configuration, applying ReLU activations and a softplus output layer to restrict the fitted bridge to positive values. Both models are optimized via Adam; $\hat{d}$ and $\hat{q}$ are trained for 200 epochs with a learning rate of $5 \times 10^{-3}$. When fitting $\hat q$, we added $0.05\sum_j\theta_j^2$ to the batch-normalized kernel loss, summing over all trainable weights and biases.

\section{Proof of Theorem \ref{thm:CPTAL}}
For $d(w,x)$,
\begin{align*}
	&\E\left[ \E\{d(W,X)|U,X,S=0\} | Z,X,S=0 \right] \\
	&= \E\left[ \E\{d(W,X)|U,Z,X,S=0\} | Z,X,S=0 \right]   \\
	&= \E\left[ d(W,X)| Z,X,S=0 \right] 
	 = \E[Y_0 - Y_\pre | Z,X,S=0] \\
	&= \E\left[ \E\{Y_0 - Y_\pre|U,Z,X,S=0\} | Z,X,S=0 \right] \\
	&= \E\left[ \E\{Y_0^{0,(0,0)} - Y_\pre^{0,(0)} | U,X,S=0\} | Z,X,S=0 \right]
\end{align*}
By Assumption \ref{A:completeness}(i), $\E[ Y_0^{0,(0,0)} - Y_\pre^{0,(0)} - d(W,X)|U,X,S=0 ]=0$.

For $q(z,x)$, we have ${p(S=1|W,X)}/{p(S=0|W,X)} = \E[ q(Z,X) | W,X,S=0 ]$.
Thus,
\begin{align*}
	&\E\left[ \E\{q(Z,X) | U,X,S=0 \} | W,X,S=0 \right] \\
	&= \E\left[ \E\{q(Z,X) | U,W,X,S=0 \} | W,X,S=0 \right] \\
	&= \E\left[ q(Z,X) | W,X,S=0 \right] 
	 = \frac{p(S=1|W,X)}{p(S=0|W,X)} \\
	&= \E\left [\frac{p(S=1|W,X)}{p(S=0|W,X)} | W,X,S=1 \right] \\
	&= \E\left[ \frac{p(S=1|U,W,X)}{p(S=0|U,W,X)} \Big| W,X,S=0 \right] \\
	&= \E\left[ \frac{p(S=1|U,X)}{p(S=0|U,X)} \Big| W,X,S=0 \right].
\end{align*}
By Assumptions \ref{A:positivity} and \ref{A:completeness}(ii), $\E[ q(Z,X) | U,X,S=0 ] = p(S=1|U,X) / p(S=0|U,X)$.

Suppose $d$ and $d'$ are solutions of Equation \eqref{eq:d}. By Assumption \ref{A:completeness_add}, $d=d'$ and the uniqueness is shown. By the same logic, $q$ is unique.
$\qed$

\section{Proof of Theorem \ref{thm:pid}}
Next, we show that the POR and PIPW identification formulas recover $\tau$ under $\mathcal{M}_{d}$ and $\mathcal{M}_{q}$, respectively.
\begin{align*}
	&\E[Y_\pre + d(W,X)|S=1] \\
	&= \E[ Y_\pre + \E\{d(W,X) | U,X,S=1\} | S=1] \\
	&= \E[ Y_\pre^{1,(0)} + \E\{Y_0^{0,(0,0)} - Y_\pre^{0,(0)} | U,X,S=1\} | S=1] \\
	&= \E[ Y_\pre^{1,(0)} + Y_0^{0,(0,0)} - Y_\pre^{0,(0)} | S=1] = \tau.
\end{align*}  
The second equality follows from Assumption \ref{A:ignorability} and Theorem \ref{thm:CPTAL}, and the last equality follows from Assumption \ref{A:CPTAL}. Similarly,
\begin{align*}
	&\E[ SY_\pre + (1-S) q(Z,X)(Y_0 - Y_\pre)] / p(S=1) \\
	&= \E[Y_\pre | S=1] + \E[ \E\{(1-S) q(Z,X)(Y_0 - Y_\pre) | U,X \} ] / p(S=1) \\
	&= \E[Y_\pre | S=1] + \E[ \E\{ q(Z,X)(Y_0 - Y_\pre) | U,X,S=0 \} p(S=0|U,X) ] / p(S=1) \\
	&= \E[Y_\pre | S=1] + \E[ \E\{ Y_0 - Y_\pre | U,X,S=0 \} p(S=1|U,X) ] / p(S=1) \\
	&= \E[Y_\pre^{1,(0)} | S=1] + \E[Y_0^{0,(0,0)} - Y_\pre^{0,(0)} | S=1] = \tau.
\end{align*}
The third equality follows from Theorem \ref{thm:CPTAL}. The penultimate equality follows from Assumption \ref{A:ignorability}, and the last equality follows from Assumption \ref{A:CPTAL}.

\section{Proof of Theorem \ref{thm:estimator}}
\subsection{Regularity Conditions}
We consider a parametric submodel ${P}_k$ of the semiparametric model, where ${P}_0={P}$ is the true law. We use the subscript $k$ to denote functionals evaluated under this submodel.

Following \citet{cui2024semiparametric} and \citet{robins1994estimation}, we assume the following conditions. Let $\Psi(\btheta) = (\psi, \bpsi_{\bbeta}^\top)$ and ${\btheta}^\top=(\tau, {\bbeta}^\top)$ where $\beta$ is a vector of all nuisance parameters, and $\bpsi_{\bbeta}$ is the vector of estimating equations for $\bbeta$. Let ${\btheta}_0$ be the solution of the population-level estimating equation $\E[\Psi(\btheta)]=0$. Denote $\|\cdot\|_F$ by the Frobenius norm, and $\|\cdot\|$ by the $L_2$-norm.
\begin{enumerate}
	\item $\E[\Psi(\btheta)] \ne 0$ if $\btheta \ne {\btheta}_0$.
	\item $\Var(\Psi(\btheta_0))$ is finite and positive definite.
	\item $\E[\partial\Psi(\btheta_0)/\partial{\btheta}^\top]$ exists and is invertible.
	\item A neighborhood $\Theta_0$ of $\btheta_0$ exists such that $\E[\sup_{\btheta\in \Theta_0} \|\Psi(\btheta)\|_F]$, $\E[\sup_{\btheta\in \Theta_0} \|\partial\Psi(\btheta)/\partial\btheta^\top\|_F]$, and $\E[\sup_{\btheta\in \Theta_0} \|\Psi(\btheta)\Psi(\btheta)^\top\|_F]$ are all finite.
	\item For all $\btheta^* \in \Theta_0$, $\E_{\btheta^*}[\|\Psi(\btheta^*)\|_F]$ and $\E_{\btheta^*}[\sup_{\btheta\in \Theta_0}\|\Psi(\btheta)\Psi(\btheta)^\top\|_F]$ are bounded, where $\E_{\btheta^*}[\cdot]$ refers to expectation with respect to the density given $\btheta^*$.
\end{enumerate}
These conditions are analogous to the regularity conditions in M-estimation \citep{boos2013essential} with $\Psi(\btheta)$ behaving essentially like estimating equations, and thus $\E[\Psi\Psi^\top]$ and $\E[\partial\Psi/\partial\btheta^\top]$ behaving like a generalized information matrix.

\subsection{Proof}
Under the parametric submodel, differentiating both sides of \eqref{eq:d} at $k=0$ results in
\begin{align}
&\E[\delta \cdot s(W,Y_0,Y_\pre|Z,X,S=0) | Z,X,S=0] = \E\left[ \frac{\partial}{\partial k} d_{k}(W,X)\Big|_{k=0} \bigg| Z,X,S=0 \right].  \label{fact3}
\end{align}
From the POR identification, the directional derivative of $\tau_{k}=\tau(\mathcal P_k)$ at $k=0$ is
\begin{align*}
	\frac{\partial}{\partial k} \tau_{k}\Big|_{k=0}
	&= \E[ \{  Y_{\pre} + d(W,X) \} s(W,Y_\pre,X|S=1) | S=1 ]  + \E\left[ \frac{\partial}{\partial k} d_k(W,X) \Big|_{k=0} \bigg| S=1 \right].
\end{align*}
Write the first term as
\begin{align*}
  &\E[ \{  Y_{\pre} + d(W,X) - \tau \} s(W,Y_\pre,X|S=1) | S=1 ] \\
  &= \E[ S \{  Y_{\pre} + d(W,X) - \tau \} s(W,Y_\pre,X|S=1) ] / p(S=1) \\
  &= \E[ S\{  Y_{\pre} + d(W,X) - \tau \} s(O) ] / p(S=1)
\end{align*}
and the second term as 
\begin{align*}
  &\E\left[ \frac{\partial}{\partial k} d_k(W,X) \Big|_{k=0} \frac{p(X,W|S=1)}{p(X,W)} \right] \\
  &=\E\left[ \frac{\partial}{\partial k} d_k(W,X) \Big|_{k=0} \frac{p(W,X|S=1)}{p(W,X)} \right] \\
  &= \E\bigg[ \frac{\partial}{\partial k} d_k(W,X) \Big|_{k=0} \E\{q(Z,X) | W,X,S=0\} p(S=0|W,X) \bigg] / p(S=1) \\
  &= \E\left[ \frac{\partial}{\partial k} d_k(W,X) \Big|_{k=0} \E\{(1-S) q(Z,X) | W,X\} \right] / p(S=1) \\
  &= \E\left[ (1-S) q(Z,X) \frac{\partial}{\partial k} d_k(W,X) \Big|_{k=0} \right] / p(S=1) \\
  &= \E\left[ q(Z,X) \frac{\partial}{\partial k} d_k(W,X) \Big|_{k=0} \bigg| S=0 \right] \frac{p(S=0)}{p(S=1)} \\
  &= \E\bigg( q(Z,X) \E\left[ \frac{\partial}{\partial k} d_k(W,X) \Big|_{k=0} \bigg| Z,X,S=0 \right] 
      \bigg| S=0 \bigg) \frac{p(S=0)}{p(S=1)} \\
  &= \E\bigg( q(Z,X) \E\left[ \delta \cdot s(W,Y_0,Y_\pre|Z,X,S=0) \big| Z,X,S=0 \right]  \Big| S=0 \bigg) \frac{p(S=0)}{p(S=1)} \\
  &= \E\left[ (1-S) q(Z,X) \delta \cdot s(W,Y_0,Y_\pre|Z,X,S) \right] / p(S=1) \\
  &= \E\left[ (1-S) q(Z,X) \delta \cdot s(O)  \right] / p(S=1).
\end{align*}

Combining two terms, we have the directional derivative
\begin{align*}
	\frac{\partial}{\partial k} \tau_{k}\Big|_{k=0}
	&= \E\Big( \big[ S\{ Y_{\pre} + d(W,X) - \tau \} + (1-S) q(Z,X) \delta \big] s(O)  \Big) / p(S=1).
\end{align*}
Hence, we claim that $\psi$ is the efficient influence function of $\tau$. It suffices to show that $\psi \in \Lambda_1 \oplus \Lambda_2$, where
\begin{align*}
	&\Lambda_1 := \{  s(Z,X,S) \in L_2(Z,X,S)\!:~ \E[s(Z,X,S)]=0  \}, \\
	&\Lambda_2 := \{  s(W,Y_0,Y_\pre|Z,X,S) \in L_2(Z,X,S)^\perp\!:~  \E[\delta \, s(W,Y_0,Y_\pre|Z,X,S=0)|Z,X,S=0] \in \overline{R(F)} \}.
\end{align*}
Notice that
\begin{align}
\psi
  &= \E\big[ S\{ Y_{\pre} + d(W,X) - \tau \} \big| Z,X,S \big] / p(S=1)  \label{eq:inlam1cc} \\
  &\quad+  S\{Y_{\pre} + d(W,X) - \tau \} / p(S=1) \\
  &\quad- \E\big[  S\{Y_{\pre} + d(W,X) - \tau \} \big| Z,X,S \big] / p(S=1)  \\
  &\quad+ [(1-S) q(Z,X) \delta ]/p(S=1)
\end{align}
The first term \eqref{eq:inlam1cc} is in $\Lambda_1$, and the sum of the remaining terms is in $\Lambda_2$ because
\begin{gather*}
	\E[ (1-S) q(Z,X) \delta | Z,X,S ] = 0, \qquad
	\E[ q(Z,X) \delta^2 | Z,X,S=0 ] \in \overline{R(F)}.
\end{gather*}

Since $d^*=d$ under $\mathcal M_d$ and $q^*=q$ under $\mathcal M_q$, under $\mathcal M_d$,
\begin{align*}
&\E[ (1-S) q^*(Z,X) \delta] = \E[ q^*(Z,X) \E\{\delta|Z,X,S=0\} | S=0]  p(S=0) = 0.
\end{align*}

Under $\mathcal M_q$, we observe that
\begin{align*}
&\E[ d^*(W,X) \{ S - (1-S) q(Z,X) \} ] = \E[ d^*(W,X) \E\{ S - (1-S) q(Z,X) | W,X \} ] = 0.
\end{align*}
Hence, $\hat\tau^{\textrm{PDR}}$ is consistent for $\tau$ under either case.

Now, assume the model $\mathcal M_d \cap \mathcal M_q$ and define $\varphi = \psi + \tau$ so that $\hat\tau^{\textrm{PDR}} = \pr_n\{\hat\varphi\}$. We have
\begin{align*}
&\sqrt n(\hat\tau^{\textrm{PDR}} - \tau) 
   = \sqrt{n}(\pr_n-\pr)\{\varphi\}
     + \sqrt{n}(\pr_n-\pr)\{\hat\varphi - \varphi\}
     + \sqrt{n}\pr\{\hat\varphi - \varphi\}.
\end{align*}
Again, the first term converges in distribution to $\mathcal N(0, \E[\psi^2])$ as $n\to\infty$, and the second term is $o_P(1)$ by Lemma 2 in \citet{kennedy2020sharp}. The last term is $o_P(1)$ since 
\begin{align*}
	|\pr\{\hat\varphi - \varphi\}| 
 	&\le \left[K_1 \|F'(\hat{q}-q)\| \|\hat{d}-d\| \right] \wedge \left[K_2 \|\hat{q}-q\| \|F(\hat{d}-d)\| \right]
\end{align*}
for finite constants $K_1$, $K_2$. Hence, $\sqrt n(\hat\tau^{\textrm{PDR}} - \tau)$ is asymptotically normal with mean $0$ and variance $\E[\psi^2]$.
$\qed$

\section{Sensitivity Analysis}  \label{sec:sensitivity}
Using the configuration in Section \ref{sec:sim} as a baseline, we introduced violations of proxy conditions by adding a direct association between variables. We evaluated levels of violations: in the ``mild'' scenario, the association that breaks the proxy condition is 10\% as strong as the true association with $U$; in the ``moderate'' scenario, this strength is increased to approximately 50\%. Details for each setting are provided below.

\paragraph{1. Violation of $Z \perp Y_0 - Y_{\pre} \mid (X, U, S)$.}
We modified the linear factor model from Section \ref{sec:sim} by incorporating a direct effect of $Z$ on the outcome trend. Specifically, we added the term $\gamma_z Z_{\text{std}}$ to the outcome model, where $Z_{\text{std}}$ represents the standardized version of $Z$. Levels of violation correspond to $\gamma_z = 0.3$ (mild) and $\gamma_z = 1.5$ (moderate).

\paragraph{2. Violation of $W \perp S \mid (X, U)$.}
Based on the logistic model for $S$ described in Section \ref{sec:sim}, we induced a direct association by incorporating the term $\gamma_w W_{\text{std}}$, where $W_{\text{std}}$ denotes the standardized version of $W$. Levels of violation correspond to $\gamma_w = 0.05$ (mild) and $\gamma_w = 0.25$ (moderate).

\paragraph{3. Violation of CPTL.} 
We propose a sensitivity analysis by introducing a controlled violation of CPTL. We modified the linear factor model in Section \ref{sec:sim} such that the trend coefficients $b_t(X, U)$ are no longer stable across the extension period. Specifically, we introduced an acceleration factor $\alpha$, where coefficients are multiplied by $(1 + \alpha)$ at each subsequent time point. At $\alpha = 0.5$, CPTL is violated as the latent confounding effect accelerates rather than remaining parallel.

\begin{table}[h!]
\caption{Sensitivity analysis for violation of proxy conditions and CPTL. Bold indicates the least biased, and underline indicates the most biased method for each scenario.}
\centering
\begin{small}
\begin{sc}
\begin{tabular}{lcccccc}
\toprule
{Violation Level} & {PDR} & {POR} & {PIPW} & {MB} & {SDID} & {MC-NNM} \\ 
\midrule
$\gamma_z = 0.3$ & \textbf{-0.254} & -0.457 & -0.254 & \underline{-1.104} & -0.567 & -0.738 \\ 
$\gamma_z = 1.5$ & -0.866 & -1.090 & \textbf{-0.848} & -1.846 & -0.994 & \underline{-1.856} \\ 
$\gamma_w = 0.05$ & \textbf{-0.142} & -0.307 & -0.155 & -0.941 & -0.395 & \underline{-1.589} \\ 
$\gamma_w = 0.25$ & -0.365 & -0.528 & -0.349 & -1.073 & \textbf{-0.120} & \underline{-1.165} \\ 
$\alpha = 0.5$ & \textbf{-0.262} & -0.502 & -0.293 & -1.343 & \underline{-2.096} & -1.475 \\ 
\bottomrule
\end{tabular}
\end{sc}
\end{small}
\end{table}

In summary, as expected, we observed an increase in bias as the severity of the violation intensified. Based on these empirical results, we suggest that Violation 1 is more detrimental than 2; by contaminating both the outcome trend and the proxy $Z$, it invalidates the conditions for POR and PIPW, thereby compromising the PDR estimator. Nevertheless, in nearly all scenarios (with the exception of the moderate case in Violation 2), our proposed methods performed similarly to or better than existing baselines.

While our methods exhibit some bias under CPTL violation, under the simulated CPTL violation, the proposed estimators had smaller absolute bias than the benchmarks. We explain this discrepancy with our framework targeting the trend. The proposed method's bias is restricted to the incremental divergence from that path. Benchmarks, conversely, are sensitive to the entire cumulative difference in trends.

\section{Falsification Test of CPTL Validity}  \label{sec:cptl_test}
To further support the validity of our framework, we introduce a statistical test that can be implemented whenever multiple trial-period outcomes (e.g., $Y_{\pre_1}, Y_{\pre_2}$) are available. This is a common scenario in regulatory settings, where short-term RCTs typically record at least three distinct time points. Intuitively, this tests whether EC and RCT control trends are parallel during the trial period.

We expand CPTL to the earlier measurement times used by the test:
$$
\E[Y_t^{1,(0)}-Y_t^{0,(0)}\mid X,U]=b(X,U),
\qquad t\in\{\pre_1,\pre_2\},
$$
with the same \(b(X,U)\) as in Assumption \ref{A:CPTAL}. This implies equal conditional untreated trends between \(\pre_1\) and \(\pre_2\).
We define a trial-period trend-confounding bridge function $d_{\pre}(W, X, S)$ that models the trend between two trial-period time points in study cohort $S$ as the solution to the equation:
$$
\E[Y_{\pre_2} - Y_{\pre_1} - d_{\pre}(W, X, S) \mid U, X, S] = 0,
$$
which we estimate using the moment condition
$$
\E[Y_{\pre_2} - Y_{\pre_1} - d_{\pre}(W, X, S) \mid Z, X, S] = 0.
$$

Under CPTL, the conditional mean trend should be invariant across study cohorts ($S=0, 1$). Specifically, CPTL holds only if the difference in the trend-confounding bridge functions is zero in expectation:
\begin{align}
	\E[d_{\pre}(W, X, 1) - d_{\pre}(W, X, 0) \mid Z, X] = 0.
	\label{eq:cptl_neccsuff}
\end{align}

Although this only depends on observables, it is not straightforward to test in practice. We instead implement a test of the following necessary condition using observable data. A rejection of $H_0$ indicates that the trends in the EC and RCT control deviate significantly, implying a violation of the extended CPTL assumption.
\begin{align}
	H_0: \E\left[ \{d_{\pre}(W, X, 1) - d_{\pre}(W, X, 0)\}({\bf 1}, Z, X)^\top \right] = \mathbf{0}.
	\label{eq:cptl_necc}
\end{align}

This diagnostic uses $Z$ in the RCT, even though the estimator can be computed with $Z$ measured only in the EC.
To formally test the necessary condition for the CPTL assumption, we implement a bootstrap-based Wald-type test. Let the sample size be $N$. The procedure begins by estimating the trial-period bridge function  $\hat{d}_{\pre}$, which is done by the same process as described in Appendix \ref{sec:learning}.
Next, we construct an instrumental variable vector for each observation, $V_i = (1, Z_i, X_i)^\top$, where $k$ represents the dimension of this vector. Using these estimated bridge functions and the instruments, we calculate the empirical moment vector for the original sample. This vector, $\bar{m}$, is defined as the sample average of the differences between the bridge functions, weighted by the instrument vector:
$$
  \bar{m} = \frac{1}{N} \sum_{i=1}^N \left( \hat{d}_{\pre}(W_i, X_i,1) - \hat{d}_{\pre}(W_i, X_i,0) \right) V_i.
$$

To estimate the variance of these moments, we employ a non-parametric bootstrap. For each bootstrap iteration $b \in \{1, \dots, B\}$, we draw a sample of size $N$ with replacement from the original dataset. We re-estimate the bridge function to obtain $\hat{d}_{\pre}^{(b)}$ based solely on this sample. We then compute the corresponding bootstrapped empirical moment vector, $\bar{m}^{(b)}$, using the resampled data and the newly estimated bridge functions.

Upon completing the $B$ bootstrap iterations, we construct the empirical covariance matrix of the moments, denoted by $\hat{V}_{boot}$. This matrix is calculated as the sample covariance of the $B$ bootstrapped moment vectors, $\{ \bar{m}^{(b)} \}_{b=1}^B$. Finally, we formulate the Wald test statistic to evaluate the null hypothesis that the moment conditions are satisfied. To ensure numerical stability, particularly if the variance-covariance matrix is ill-conditioned or singular, we utilize the Moore-Penrose pseudoinverse, $\hat{V}_{boot}^{+}$. The Wald statistic is thus calculated as $W_{stat} = \bar{m}^\top \hat{V}_{boot}^{+} \bar{m}.$

Under the null hypothesis, this test statistic asymptotically follows a chi-square distribution with $k$ degrees of freedom, where $k$ corresponds to the length of the instrument vector $V_i$. Consequently, the statistical significance of the test is determined by calculating the $p$-value, given by $1 - F_{\chi^2_k}(W_{stat})$, where $F_{\chi^2_k}$ represents the cumulative distribution function of the $\chi^2_k$ distribution.
    
To evaluate its practical utility, we applied this test to the accelerated linear factor model described in the previous section. In a scenario with a known violation ($\alpha = 1$), the test yielded a $p$-value of $0.002$ (test statistic: $16.941$), successfully identifying the departure from CPTL. Conversely, under the null hypothesis of valid CPTL ($\alpha = 0$), the test produced a $p$-value of $0.632$ (test statistic: $2.571$), correctly failing to reject the assumption. 
Furthermore, across independently generated simulation replicates with a sample size of $n = 2500$, the test maintained a nominal Type I error rate of $0.05$ and achieved a statistical power of $0.75$.

It is important to emphasize that failing to reject the null hypothesis does not confirm that CPTL holds. Specifically, the test cannot detect a violation if the trends remain parallel during the trial period but diverge during the extension period. Furthermore, the test is insensitive to scenarios where the necessary condition in \eqref{eq:cptl_necc} holds but \eqref{eq:cptl_neccsuff} fails. This latter limitation can be addressed by employing a more stringent version of the test that utilizes transformations of $({\bf 1},Z,X)^\top$, rather than relying solely on the untransformed covariates.

\section{\tred{Proximal Difference-in-Differences with Missing Outcomes}}\label{AA:marl}
\subsection{Missing at Random Given Latent Factors}
In practice, outcomes are subject to missingness, including dropout. Standard methods include imputation followed by complete-case analysis or methods that assume missing at random (MAR), given a sufficient set of observed covariates. These methods might introduce bias into estimation due to MAR not being met. Compounding the issue, MAR is untestable.

\begin{figure}[hbt]
	\centering
    \begin{minipage}{.2\textwidth}
	\scalebox{1}{\begin{tikzpicture}
    \node[state] (x) at (0,0) {$X$};
    \node[state] (y) [right =of x] {$Y_t$};
    \node[state] (r) [above =of y] {$R_t$};

    \path (x) edge (y);
    \path (x) edge (r);
	\end{tikzpicture}}
	
	\centering
	\text{a.}
\end{minipage}
\qquad
\begin{minipage}{.2\textwidth}
	\scalebox{1}{\begin{tikzpicture}
    \node[state] (x) at (0,0) {$X$};
    \node[state] (u) at (2.25,.95) {\footnotesize $U$};
    \node[state] (y) at (1.5,0) {$Y_t$};
    \node[state] (r) [above =of y] {$R_t$};

    \path (x) edge (y);
    \path (x) edge (r);
    \path (u) edge (r);
    \path (u) edge (y);
	\end{tikzpicture}}
	
	\centering
	\text{b.}
\end{minipage}
\qquad
\begin{minipage}{.2\textwidth}
	\scalebox{1}{\begin{tikzpicture}
    \node[state] (x) at (0,0) {$X$};
    \node[state] (y) [right =of x] {$Y_t$};
    \node[state] (r) [above =of y] {$R_t$};

    \path (x) edge (y);
    \path (x) edge (r);
    \path[bidirected] (r) edge[bend left=20] (y);
	\end{tikzpicture}}
	
	\centering
	\text{c.}
\end{minipage}
	
  \caption{Causal directed acyclic graphs illustrating missing data mechanisms: a. Missing at Random (MAR); b. Missing at Random given Latent factors (MARL); and c. Missing Not At Random (MNAR). Graphs are simplified to emphasize the relationship.}
  \label{fig:MARL}
\end{figure}
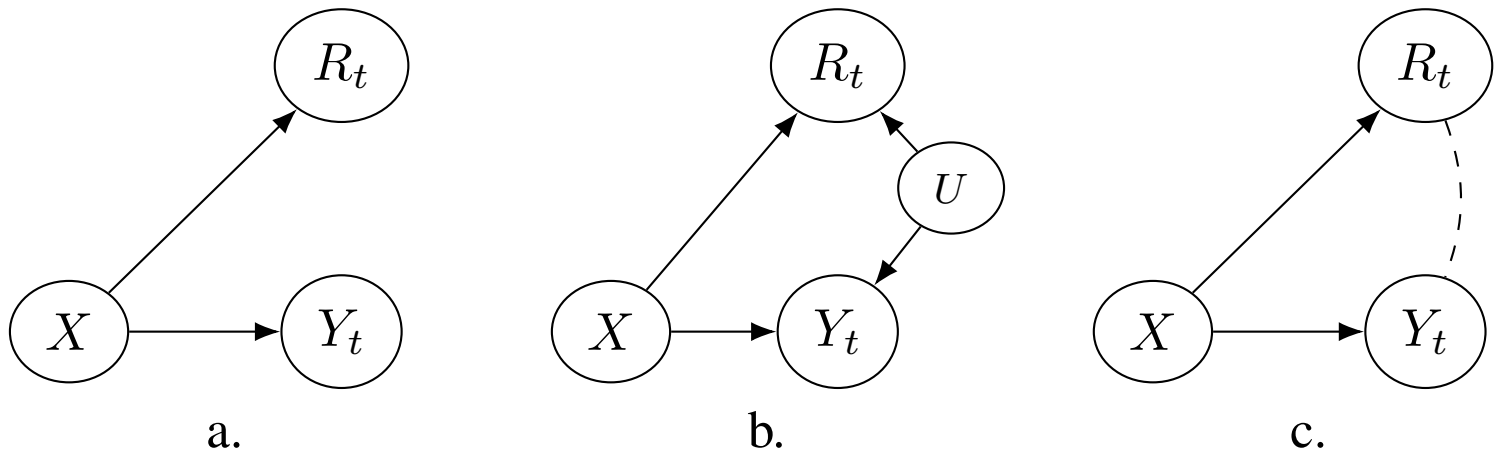

We introduce {\it Missing At Random given Latent factors (MARL)}, a condition to model the complex patterns of missing data observed in real-world scenarios.   We define $R_0$ and $R_\pre$ to be missingness indicators of $Y_0$ and $Y_\pre$, respectively, meaning that $R_0=1$ or $R_\pre=1$ if $Y_0$ or $Y_\pre$ are observed, respectively, and $R_0=0$ or $R_\pre=0$ otherwise. Let $\bar{R} = (R_\pre, R_0)$.
\tred{
\begin{assumption} \label{A:proxy_cond_marl}
	$(Y_0, Y_\pre, \bar R) \ind Z \mid U,X,S$ and $W \ind (S, Z, \bar R) \mid U, X$.
\end{assumption} 
Assumption \ref{A:proxy_cond_marl} extends the proxy conditions to account for outcome missingness. Conditional on the stated factors, neither proxy provides additional information about whether outcomes are observed. The proxies may nevertheless be associated with missingness through $U$ and $X$. This excludes, for example, a direct effect of $Z$ on follow-up attendance or of $W$ on dropout beyond these factors.
}
\begin{assumption} \label{A:MARL} \
(i) $(R_0^{(0,0)},R_\pre^{(0)}) \ind W \mid U,X$.
\;(ii) $(R_0^{(0,0)},R_\pre^{(0)}) \ind (Y_0^{(0,0)}, Y_\pre^{(0)}) \mid U,X,S$.
\end{assumption}
Assumption \ref{A:MARL}(i) posits the existence of a proxy variable $W$ that is independent of missingness given the confounders.
Under MAR, missingness of a variable is independent of unobserved values, conditional on the observed variables. Part (ii) instead assumes that missingness in the outcomes can be fully explained by observed \textit{and} latent variables. Consequently, MARL is a special case of missing not at random (MNAR) that allows for identifiability through the use of proxy variables (Figure \ref{fig:MARL}). 
MARL can model the patterns that frequently occur in real-world data, including dropout, i.e., $R_0 \le R_\pre$.

\subsection{Bridge Functions}
\begin{assumption}[Positivity of study participation] \label{A:positivity_marl} \
	$0<\pr(S=1, R_\pre=1| X,U)<1$ and $0<\pr(S=0, R_0=R_\pre=1| X,U)<1$ almost surely.
\end{assumption}
Assumption \ref{A:positivity_marl} is analogous to Assumption \ref{A:positivity} in the fully observed study, which states that for each study cohort, the probability that a subject is observed and originates from that cohort must be positive.

\begin{assumption}[Completeness] \label{A:completeness_marl} \
For a square-integrable $v$,
\begin{enumerate}
	\item[] (i) $\E\left[ v(U) | z,x,s,R_\pre=1 \right] = 0$ for all $z,x$ and $s$ only if $v(U)=0$.
	\item[] (ii) $\E\left[ v(U) | z,x,S=0,R_0=1 \right] = 0$ for all $z$ and $x$ only if $v(U)=0$.
	\item[] (iii) $\E\left[ v(U) | w,x,S=0,R_0=1 \right] = 0$ for all $w$ and $x$ only if $v(U)=0$.
	\item[] (iv) $\E\left[ v(U) | w,x,S=s, R_\pre=1 \right] = 0$, $s\in\{0,1\}$, for all $w,x$ only if $v(U)=0$.
\end{enumerate}
\end{assumption}
Assumption \ref{A:completeness_marl} is analogous to Assumption \ref{A:completeness}, which states that the proxies should have enough variability compared to the variability of $U$, specifically in the observed subpopulation.
Bridge functions are now defined on each observed subpopulation of study cohorts.

\begin{definition}[Bridge functions] \label{def:bridge_CPTAL_marl} \
Define bridge functions $h_0(w,x)$, $h_\pre(w,x,s)$, $q_0(z,x)$ and $q_\pre(z,x,s)$ as the solutions of the following integral equations for $s=0,1$:
	\begin{align}
		\E[Y_\pre-h_\pre(W,X,S)|Z,X,S,A_\pre=0,R_\pre=1] = 0,  \label{eq:hpre} \\
		\E[Y_0 - h_0(W,X)|Z,X,S=0,R_0=1] = 0,  \label{eq:h} \\
		\E[(1-S)R_0 q_0(Z,X) - S | W,X] = 0,  \label{eq:q}  \\
		\E[\mathbf1\{S=s\}R_\pre q_\pre(Z,X,s) - S | W,X] = 0.  \label{eq:eta}
	\end{align}
\end{definition}

\tred{Note that in \eqref{eq:eta}, the inner term subtracts $S$, rather than $\mathbf{1}\{S=s\}$, for both $s=0$ and $s=1$.}
Here, $h_\pre(w,x,1)-h_\pre(w,x,0)$ connects to the conditional bias $b(x,u)$, $h_0(w,x)$ connects to the EC mean $\mu_t(x,u)$, and $(q_0, q_\pre)$ connect to inverse probability weights.
These links between bridge functions and functionals conditional on latent factors are formally established by the following theorem.

\begin{theorem}  \label{thm:CPTAL_marl}
Under Assumptions \ref{A:CPTAL}, \ref{A:ignorability}, \ref{A:proxy_cond_marl}-\ref{A:completeness_marl}, and \ref{A:Acompleteness_marl}, $h_0$, $h_\pre$, $q_0$, $q_\pre$ are unique solutions of
	\begin{align*}
		&\E[Y_\pre^{S,(0)}-h_\pre(W,X,S) | U,X,S]=0,\\
		&\E[Y_0^{0,(0,0)} - h_0(W,X) | U,X,S=0]=0,\\
		&\E[ q_0(Z,X) | U,X,S=0,R_0=1 ] = \frac{p(S=1|U,X)}{p(S=0,R_0=1|U,X)},  \\
		&\E[ q_\pre(Z,X,s) | U,X,S=s,R_\pre=1 ] = \frac{p(S=1|U,X)}{p(S=s,R_\pre=1|U,X)}, \quad s=0,1.
	\end{align*}
\end{theorem}
\begin{proof}
	$\E[ Y_0^{0,(0,0)}-h_0(W,X)|U,X,S=0 ]=0$ follows from Assumption \ref{A:completeness_marl}(ii) and
\begin{align*}
	&\E\left[ \E\{h_0(W,X)|U,X,S=0\} | Z,X,S=0,R_0=1 \right] \\
	&= \E\left[ \E\{h_0(W,X)|U,Z,X,S=0,R_0=1\} | Z,X,S=0,R_0=1 \right]  \tag{$\because$ Assumptions \ref{A:proxy_cond_marl}, \ref{A:MARL}(i)} \\
	&= \E\left[ h_0(W,X)| Z,X,S=0,R_0=1 \right] 
	 = \E[Y_0|Z,X,S=0,R_0=1] \\
	&= \E\left[ \E\{Y_0|U,Z,X,S=0,R_0=1\} | Z,X,S=0,R_0=1 \right] \\
	&= \E\left[ \E\{Y_0^{0,(0,0)}|U,X,S=0\} | Z,X,S=0,R_0=1 \right].  \tag{$\because$ Assumption \ref{A:MARL}(ii)}
\end{align*}
Similarly, $\E[ Y_\pre^{S,(0)}-h_\pre(W,X,S) | U,X,S ]=0$ by Assumption \ref{A:completeness_marl}(i).
For $q_0(z,x)$ and $q_\pre(z,x,s)$, 
\begin{align*}
	&\E\left[ \E\{q_0(Z,X) | U,X,S=0,R_0=1 \} | W,X,S=0,R_0=1 \right] \\
	&= \E\left[ \E\{q_0(Z,X) | U,W,X,S=0,R_0=1 \} | W,X,S=0,R_0=1 \right] \\
	&= \E\left[ q_0(Z,X) | W,X,S=0,R_0=1 \right] = \frac{p(S=1|W,X)}{p(S=0,R_0=1|W,X)} \\
	&= \E\left [\frac{p(S=1|W,X)}{p(S=0,R_0=1|W,X)} | W,X,S=1 \right] \\
	&= \E\left[ \frac{p(S=1|U,W,X)}{p(S=0,R_0=1|U,W,X)} \Big| W,X,S=0,R_0=1 \right] \\
	&= \E\left[ \frac{p(S=1|U,X)}{p(S=0,R_0=1|U,X)} \Big| W,X,S=0,R_0=1 \right].
\end{align*}
Finally, a similar argument gives
\begin{align*}
	&\E\left[ \E\{q_\pre(Z,X,S) | U,X,S,R_\pre=1 \} | W,X,S,R_\pre=1 \right] \\
	&= \E\left[ \frac{p(S=1|U,X)}{p(S,R_\pre=1|U,X)} \Big| W,X,S,R_\pre=1 \right].
\end{align*}
By Assumptions \ref{A:positivity_marl} and \ref{A:completeness_marl}, the desired result follows.
\end{proof}

Similar to Appendix \ref{A:bridge_existence}, the existence of solutions to \eqref{eq:hpre}-\eqref{eq:eta} follows immediately from Picard's theorem (\cite[Theorem 15.18]{kress1989linear}) under corresponding regularity and completeness conditions. To elaborate, let $L^2(p(t))$ denote the space of square integrable functions with respect to the distribution $p(t)$, and define an inner product $\langle f, g \rangle = \int f(t)g(t)p(t)dt$ of this Hilbert space.
Let $F_{x}: L^2(p(w|x,S=0,R_0=1)) \to L^2(p(z|x,S=0,R_0=1))$, $F_{x}h = \E[h(W)|z,x,S=0,R_0=1]$ and $F'_{x}: L^2(p(z|x,S=0,R_0=1)) \to L^2(p(w|x,S=0,R_0=1))$, $F'_{x}q = \E[q(Z)|w,x,S=0,R_0=1]$. 
Similarly, let $F_{\pre,x,s}: L^2(p(w|x,s,R_\pre=1)) \to L^2(p(z|x,s,R_\pre=1))$, $F_{\pre,x,s}h = \E[h(W)|z,x,s,R_\pre=1]$ and $F'_{\pre,x,s}: L^2(p(z|x,s,R_\pre=1)) \to L^2(p(w|x,s,R_\pre=1))$, $F'_{\pre,x,s}q_\pre = \E[q_\pre(Z)|w,x,s,R_\pre=1]$.
Let $(\lambda_{x,n}, \varphi_{x,n}, \phi_{x,n})_{n=1}^\infty$, $(\lambda_{\pre,x,s,n}, \varphi_{\pre,x,s,n}, \phi_{\pre,x,s,n})_{n=1}^\infty$, $(\lambda'_{x,n}, \varphi'_{x,n}, \phi'_{x,n})_{n=1}^\infty$, $(\lambda'_{\pre,x,s,n}, \varphi'_{\pre,x,s,n}, \phi'_{\pre,x,s,n})_{n=1}^\infty$ be singular value decompositions of $F_{x}$, $F_{\pre,x,s}$, $F'_{x}$, and $F'_{\pre,x,s}$, respectively.

We assume regularity conditions:
\begin{enumerate}
	\item[1.] $\iint p(w|z,x,s,R_\pre=1) p(z|w,x,s,R_\pre=1) dwdz < \infty$ and \\$\iint p(w|z,x,S=0,R_0=1) p(z|w,x,S=0,R_0=1) dwdz < \infty$;
	\item[2.] $\E[Y_0|z,x,S=0,R_0=1] \in L^2(p(z|x,S=0,R_0=1))$ and \\$\E[Y_\pre|z,x,s,R_\pre=1] \in L^2(p(z|x,s,R_\pre=1))$;
	\item[2'.] ${p(S=1|W,X)}/{p(S=0,R_0=1|W,X)} \in L^2(p(w|x,S=0,R_0=1))$ and \\
	    ${p(S=1|W,X)}/{p(S=s,R_\pre=1|W,X)} \in L^2(p(w|x,s,R_\pre=1))$;
	\item[3.] $\sum_{n=1}^\infty \lambda_{x,n}^{-2}|\langle \E[Y_0|z,x,S=0,R_0=1], \phi_{x,n} \rangle|^2 < \infty,$ \\
	  $\sum_{n=1}^\infty \lambda_{\pre,x,s,n}^{-2}|\langle \E[Y_\pre|z,x,s,R_\pre=1], \phi_{\pre,x,s,n} \rangle|^2 < \infty$;
	\item[3'.] $\sum_{n=1}^\infty \lambda_{x,n}^{'-2}|\langle \frac{p(S=1|W,X)}{p(S=0,R_0=1|W,X)}, \phi'_{x,n} \rangle|^2, 
	  \sum_{n=1}^\infty \lambda_{\pre,x,s,n}^{'-2}|\langle \frac{p(S=1|W,X)}{p(S=s,R_\pre=1|W,X)}, \phi'_{\pre,x,s,n} \rangle|^2 < \infty$.
\end{enumerate}
By condition 1, $F_x$, $F'_x$, $F_{\pre,x,s}$, and $F'_{\pre,x,s}$ are compact operators for any $x,s$ and thus admit singular value decompositions.
The existence of solutions to \eqref{eq:hpre}-\eqref{eq:eta} follows immediately from Picard's theorem (\cite[Theorem 15.18]{kress1989linear}) under the following assumption:
\begin{assumption} \label{A:Acompleteness_marl} 
For a square integrable $v$,
\begin{enumerate}
    \item[(i)] $\E[v(W) | z,x,s,R_\pre=1]=0$ for all $z,x$ and $s$ only if $v(W)=0$.
    \item[(ii)] $\E[v(W) | z,x,S=0,R_0=1]=0$ for all $z,x$ only if $v(W)=0$.
    \item[(iii)] $\E[v(Z) | w,x,S=0,R_0=1]=0$ for all $w,x$ only if $v(Z)=0$.
    \item[(iv)] $\E[v(Z) | w,x,s,R_\pre=1]=0$ for all $w,x$ and $s$ only if $v(Z)=0$.
\end{enumerate}
\end{assumption}

\subsection{Proximal Identification}
Let $\mathcal M_{sp}$ be the semiparametric model that posits no restriction on the observed data distribution other than the existence of $h_0$ and $h_{\pre}$ \citep{cui2024semiparametric}.
Let $F: L_2(W,X|S=0,R_0=1) \to L_2(Z,X|S=0,R_0=1)$, $F:g \mapsto \E[g(W,X)|Z,X,S=0,R_0=1]$ and 
$F_\pre: L_2(W,X,S|R_\pre=1) \to L_2(Z,X,S|R_\pre=1)$, $F_\pre:g \mapsto \E[g(W,X,S)|Z,X,S,R_\pre=1]$ be the conditional expectation operators and
let $F': L_2(Z,X|S=0,R_0=1) \to L_2(W,X|S=0,R_0=1)$ and $F'_\pre: L_2(Z,X,S|R_\pre=1) \to L_2(W,X,S|R_\pre=1)$ 
defined by $F':g \mapsto {\E}[ g(Z, X) | W, X, S=0, R_0=1 ]$ and $F'_\pre:g \mapsto \E[g(Z,X,S)|W,X,S,R_\pre=1]$ be the adjoint operators of $F$ and $F_\pre$, respectively. 

Define proximal outcome regression (POR), proximal inverse probability weighting (PIPW), and proximal doubly robust (PDR) estimators
	\begin{align}
		\hat{\tau}^{\textrm{POR}} 
		&= \E_n[ \hat h_0(W,X) + \hat h_{\pre}(W,X,1) - \hat h_{\pre}(W,X,0) | S=1 ],  \label{eq:POR_marl}  \\ 
		\hat{\tau}^{\textrm{PIPW}} 
		&= \E_n\left[ SR_{\pre} \hat{q}_\pre(Z,X,1) Y_{\pre} + (1{-}S)\{ R_0\hat{q}_0(Z,X)Y_0 {-} R_{\pre}\hat{q}_\pre(Z,X,0)Y_{\pre} \} \right] / \pr_n(S{=}1),  \label{eq:PIPW_marl} \\ 
		\hat{\tau}^{\textrm{PDR}} 
		&= \E_n[ S\{ \hat{h}_0(W,X) + \hat{h}_{\pre}(W,X,1) - \hat{h}_{\pre}(W,X,0) + R_{\pre} \hat{q}_\pre(Z,X,1) \hat{e}_{\pre}(1) \}  \\
 		&\qquad\;+ (1-S)\{ R_0\hat{q}_0(Z,X) \hat{e}_0 - R_{\pre}\hat{q}_\pre(Z,X,0)\hat{e}_{\pre}(0)\} ] / \pr_n(S=1),  \label{eq:PDR_marl}
 	\end{align}
 	where ${e}_0 = Y_0 - h_0(W,X)$ and ${e}_\pre(s) = Y_\pre - {h}_\pre(W,X,s)$. 
\tred{Unlike the estimators in Sections \ref{sec:id} and \ref{sec:pdr}, these estimators additionally require $Z$ to be observed in the RCT.}
The next theorem shows that these estimators are consistent, and the PDR estimator additionally enjoys double robustness as well as semiparametric local efficiency.

\tred{
\begin{assumption}  \label{A:reg_marl} \
\begin{enumerate}[topsep=0pt,itemsep=-0.5ex,partopsep=1ex,parsep=1ex]
	\item[(i)] $\hat h_\pre$, $\hat h_0$, $\hat q_\pre$ and $\hat q_0$ converge in probability to $h_\pre^*$, $h_0^*$, $q_\pre^*$ and $q_0^*$, respectively.
	\item[(ii)] $\pr_n\{\hat h_\pre\}$, $\pr_n\{\hat h_0\}$, $\pr_n\{\hat q_\pre\}$, and $\pr_n\{\hat q_0\}$ converge in probability to $\pr\{h_\pre^*\}$, $\pr\{h_0^*\}$, $\pr\{q_\pre^*\}$, and $\pr\{q_0^*\}$, respectively, as $n \to \infty$.
	\item[(iii)] $\hat h_\pre$, $\hat h_0$, $\hat q_\pre$ and $\hat q_0$ are estimated via cross-fitting, i.e., the estimated law $\hat\pr$ associated with the nuisance estimators is estimated independently of $\pr_n$.
	\item[(iv)] $\|\hat h_\pre\|$, $\|\hat h_0\|$, $\|\hat q_\pre\|$, and $\|\hat q_0\|$ are uniformly bounded in the neighborhood $\Theta_0$ of the true nuisance functions.
\end{enumerate}
\end{assumption}
}
\tred{
\begin{assumption}  \label{A:rate_marl}
	For all $\btheta \in \Theta_0$, (i) $\|F'(\hat q_0 - q_0)\| \|\hat h_0 - h_0\| = o_P(n^{-1/2})$, 
	$\|F'(\hat q_0 - q_0)\| \|\hat h_\pre - h_\pre\| = o_P(n^{-1/2})$, 
	and $\|F'_\pre(\hat q_\pre - q_\pre)\| \|\hat h_\pre - h_\pre\| = o_P(n^{-1/2})$, 
	or (ii) $\|\hat q_0 - q_0\| \|F(\hat h_0 - h_0)\| = o_P(n^{-1/2})$, 
	$\|\hat q_0 - q_0\| \|F_\pre(\hat h_\pre - h_\pre)\| = o_P(n^{-1/2})$, and 
	$\|\hat q_\pre - q_\pre\| \|F_\pre(\hat h_\pre - h_\pre)\| = o_P(n^{-1/2})$.
\end{assumption}
}

\tred{
\begin{theorem}[Proximal estimators] \label{thm:estimator_marl} 		
Assume $\pr_n(S=1) \to \pr(S=1)$ as $n\to\infty$. Suppose Assumptions \ref{A:CPTAL}, \ref{A:ignorability}, and \ref{A:proxy_cond_marl}-\ref{A:reg_marl} and regularity conditions hold.
\begin{enumerate}
	\item[(i)] $\hat\tau^{\textrm{POR}}$, $\hat\tau^{\textrm{PIPW}}$ are consistent estimators of $\tau$ with correctly specified $\mathcal{H}=(h_0,h_\pre)$, $\mathcal{Q} = (q_0,q_\pre)$, respectively.
	\item[(ii)] $\hat\tau^{\textrm{PDR}}$ is a consistent estimator of $\tau$ with correctly specified $\mathcal{H}=(h_0,h_\pre)$ or $\mathcal{Q} = (q_0,q_\pre)$.
 	\item[(iii)] $
		\psi 
		= [ S\{ h_0(W,X) + h_\pre(W,X,1) - h_\pre(W,X,0) + R_\pre q_\pre(Z,X,1)e_\pre(1) - \tau \} 
		+ (1-S)\{R_0 q_0(Z,X)e_0 - R_\pre q_\pre(Z,X,0) e_\pre(0)\} ] / \pr(S=1)
	$ is the efficient influence function of $\tau$ under $\mathcal{M}_{sp}$ evaluated at the submodel that $F$, $F_\pre$, $F'$, $F'_\pre$  are surjective.
 	\item[(iv)] Suppose additionally Assumption \ref{A:rate_marl} holds. With correctly specified $\mathcal{H}$ and $\mathcal{Q}$, $\hat\tau^{\textrm{PDR}}$ is consistent for $\tau$ as $n\to\infty$, and is semiparametric locally efficient, with asymptotic variance $\E[\psi^2]$, under the submodel that $F$, $F_\pre$, $F'$, $F'_\pre$  are surjective.
\end{enumerate}
\end{theorem}
}
When all outcomes are observed, i.e., $R_\pre=R_0=1$, $q_\pre(\cdot,\cdot,1)=1$, $q_\pre(\cdot,\cdot,0)=q_0$, and $d=h_0-h_\pre(\cdot,\cdot,0)$, the PDR estimator simplifies to the doubly robust form in Section \ref{sec:pdr}.

\begin{proof} 
Again, we simplify the problem by assuming the only available data sources are external data and trial control data during the trial period; we disregard trial treated data as it does not affect estimation and inference of the trial control trajectory.
Let $O=\{Z,X,W,R_\pre,R_0,R_\pre Y_\pre,(1-S)R_0Y_0,S\}$ be the set of all observed variables. 
We use $\mathcal{M}$ with subscripts $sp$, $\mathcal{H}$, $\mathcal{Q}$ to denote a semiparametric model, a model with correct specification of the outcome confounding bridge functions, and a model with correct specification of the source-confounding bridge functions, respectively.    
Let $\pr$ be the true distribution generating the data, $p$ be the corresponding density or mass, and $s(\cdot)$ be the score function.   Let $\pr_n$ be the empirical distribution from $\{O_i: i=1,\dots,n\}$. Denote $\pr\{\cdot\}$ by the expectation operator and $\pr_n\{\cdot\}$ by the empirical mean operator, i.e. for a functional $f$, $\pr\{f\}=\E[f(O)]$ and $\pr_n\{f\}=\E_n[f(O)]$.

We consider a parametric submodel ${P}_k$ of the semiparametric model where ${P}_0={P}$ is the true law. We use the subscript $k$ to denote functionals evaluated under this submodel.
Let $\Psi(\btheta) = (\psi, \bpsi_{\bbeta}^\top)$ and $\btheta^\top=(\tau, \bbeta^\top)$ where $\beta$ is a vector of all nuisance parameters, and $\bpsi_{\bbeta}$ is the vector of estimating equations for $\bbeta$. Let $\btheta_0$ be the solution of the population-level estimating equation $\E[\Psi(\btheta)]=0$. Denote $\|\cdot\|_F$ by the Frobenius norm, and $\|\cdot\|$ by the $L_2$-norm.
Following \cite{cui2024semiparametric, robins1994estimation}, we assume the following conditions. 
\begin{enumerate}
	\item $\E[\Psi(\btheta)] \ne 0$ if $\btheta \ne \btheta_0$; \quad 2. $\Var(\Psi(\btheta_0))$ is finite and positive definite;
	\item[3.] $\E[\partial\Psi(\btheta_0)/\partial\btheta^\top]$ exists and is invertible;
	\item[4.] A neighborhood $\Theta_0$ of $\btheta_0$ exists such that $\E[\sup_{\btheta\in \Theta_0} \|\Psi(\btheta)\|_F]$, $\E[\sup_{\btheta\in \Theta_0} \|\partial\Psi(\btheta)/\partial\btheta^\top\|_F]$, and $\E[\sup_{\btheta\in \Theta_0} \|\Psi(\btheta)\Psi(\btheta)^\top\|_F]$ are all finite;
	\item[5.] For all $\btheta^* \in \Theta_0$, $\E_{\btheta^*}[\|\Psi(\btheta^*)\|_F]$ and $\E_{\btheta^*}[\sup_{\btheta\in \Theta_0}\|\Psi(\btheta)\Psi(\btheta)^\top\|_F]$ are bounded, where $\E_{\btheta^*}[\cdot]$ refers to expectation with respect to the density given $\btheta^*$.
\end{enumerate}
These conditions are analogous to the regularity conditions in M-estimation \citep{boos2013essential} with $\Psi(\btheta)$ behaving essentially like estimating equations, and thus $\E[\Psi\Psi^\top]$ and $\E[\partial\Psi/\partial\btheta^\top]$ behaving like a generalized information matrix.

Under the parametric submodel, differentiating both sides of \eqref{eq:hpre}-\eqref{eq:h} at $k=0$ gives
\begin{align}
&\E[e_\pre(S) s(W,Y_\pre|Z,X,S,R_\pre=1) | Z,X,S,R_\pre=1] \nonumber \\
&= \E\left[ \frac{\partial}{\partial k} h_{\pre,k}(W,X,S)\Big|_{k=0} \bigg| Z,X,S,R_\pre=1 \right],  \label{fact1}  \\
&\E[e_0 \cdot s(W,Y_0|Z,X,S=0,R_0=1) | Z,X,S=0,R_0=1] \nonumber \\
&= \E\left[ \frac{\partial}{\partial k} h_{0,k}(W,X)\Big|_{k=0} \bigg| Z,X,S=0,R_0=1 \right].  \label{fact2}
\end{align}
Now, from \eqref{eq:POR_marl}, the directional derivative of $\tau_{k}=\tau(\mathcal P_k)$ at $k=0$ is
\begin{align}
	\frac{\partial}{\partial k} \tau_{k}\Big|_{k=0}
	&= \E[ \{  h_0(W,X) + h_{\pre}(W,X,1) - h_{\pre}(W,X,0) \} s(W,X|S=1) | S=1 ]  \label{eq:eif1} \\
	&\quad+ \E\left[ \frac{\partial}{\partial k} \{  h_0(W,X) + h_{\pre}(W,X,1) - h_{\pre}(W,X,0) \} \Big|_{k=0} \bigg| S=1 \right].  \label{eq:eif2}
\end{align}
Organizing each term gives
\begin{align*}
\eqref{eq:eif1}	
  &= \E[ S\{  h_0(W,X) +  h_{\pre}(W,X,1) -  h_{\pre}(W,X,0) - \tau \} s(O) ] / \pr(S=1), \\
\eqref{eq:eif2}	
  &= \E\left[ \frac{\partial}{\partial k} \{  h_0(W,X) - h_{\pre}(W,X,0) \} \Big|_{k=0} \frac{p(X,W|S=1)}{p(X,W)} \right] 
  + \E\left[ \frac{\partial}{\partial k} h_{\pre}(W,X,1) \Big|_{k=0} \bigg| S=1 \right].
\end{align*}
Further organizing the first two terms using the facts in \eqref{fact1} and \eqref{fact2} leads to
\begin{align*}
&\E\left[ \frac{\partial}{\partial k} h_0(W,X) \Big|_{k=0} \frac{p(W,X|S=1)}{p(W,X)} \right] 
  = \E\left[ \frac{\partial}{\partial k} h_0(W,X) \Big|_{k=0} \E\{(1{-}S)R_0 q_0(Z,X) | W,X\} \right] / p(S=1) \\
  &= \E\bigg( q_0(Z,X) \E\left[ \frac{\partial}{\partial k}  h_0(W,X)  \Big|_{k=0} \bigg| Z,X,S=0,R_0=1 \right]  \bigg| S=0,R_0=1 \bigg) \frac{p(S=0,R_0=1)}{p(S=1)} \\
  &= \E\Big( q_0(Z,X) \E\left[  e_0 \, s(W,Y_0|Z,X,S=0,R_0=1) \big| Z,X,S=0,R_0=1 \right] \Big| S=0,R_0=1 \Big) \frac{p(S=0,R_0=1)}{p(S=1)} \\
  &= \E\left[ (1-S)R_0 q_0(Z,X) e_0 \cdot s(O)  \right] / p(S=1), \\
&\E\left[ \frac{\partial}{\partial k} h_{\pre}(W,X,0) \Big|_{k=0} \frac{p(W,X|S=1)}{p(W,X)} \right] = \E\left[ (1-S)R_\pre q_\pre(Z,X,0) e_{\pre}(0) s(O)  \right] / p(S=1).
\end{align*}
Similarly, the third term is simplified to
\begin{align*}
&\E\left[ \frac{\partial}{\partial k} h_{\pre,k}(W,X,1) \Big|_{k=0} \bigg| S=1 \right]
  = \E\left[ SR_\pre q_\pre(Z,X,1) e_\pre(1) s(O) \right] / {p(S=1)}.
\end{align*}
Combining all terms, we have the directional derivative
\begin{align*}
	\frac{\partial}{\partial k} \tau_{k}\Big|_{k=0}
	&= \E\Big( \big[ S\{h_0(W,X) + h_{\pre}(W,X,1) - h_{\pre}(W,X,0) - \tau + R_\pre q_\pre(Z,X,1) e_\pre(1) \}  \\
 	&\qquad\;+ (1-S) \{R_0 q_0(Z,X) e_0 - R_\pre q_\pre(Z,X,0) e_{\pre}(0)\} \big] s(O)  \Big) / p(S=1).
\end{align*}
Hence, we claim that $\psi$ is the efficient influence function of $\tau$. It suffices to show that the components in $\psi$ lie in the tangent space $\Lambda_1 \oplus \Lambda_2$, where
\begin{align*}
	&\Lambda_1 := \{  s(Z,X,S,R_0,R_\pre) \in L_2(Z,X,S,R_0,R_\pre):~ \E[s(Z,X,S,R_0,R_\pre)]=0  \}, \\
	&\Lambda_2 := \{s(Y_0,Y_{\pre},W|Z,X,S,R_0,R_{\pre}) \in L_2(Z,X,S,R_0,R_{\pre})^\perp: \\ 
	&\quad\qquad \E[e_0 \, s(Y_0,Y_{\pre},W|Z,X,S=0,R_0=1,R_{\pre})|Z,X,S=0,R_0=1] \in \overline{R(F)}, \\
    &\quad\qquad \E[e_{\pre}(S) s(Y_0,Y_{\pre},W|Z,X,S,R_0,R_{\pre}=1)|Z,X,S,R_{\pre}=1] \in \overline{R(F_{\pre})} \}. 
\end{align*}
Here, $R(F)$ denotes the range of $F$, and $H^\perp$ denotes the orthogonal complement of $H$. We add and subtract the same term to achieve the decomposition:
\begin{align}
\psi
  &= \E\big[ S\{h_0(W,X) + h_{\pre}(W,X,1) - h_{\pre}(W,X,0) - \tau \} \big| Z,X,S,R_0,R_\pre \big] / p(S=1)  \label{eq:inlam1} \\
  &\quad+  S\{h_0(W,X) + h_{\pre}(W,X,1) - h_{\pre}(W,X,0) - \tau \} / p(S=1) \\
  &\quad- \E\big[  S\{h_0(W,X) + h_{\pre}(W,X,1) - h_{\pre}(W,X,0) - \tau \} \big| Z,X,S,R_0,R_\pre \big] / p(S=1)  \\
  &\quad+ [SR_\pre q_\pre(Z,X,1) e_\pre(1) + (1-S) \{R_0 q_0(Z,X)e_0 - R_\pre q_\pre(Z,X,0)e_{\pre}(0)\}]/p(S=1)
\end{align}
Under the stated surjectivity of $F$ and $F_\pre$, the two conditional moments defining $\Lambda_2$ belong to their respective ranges whenever they are square-integrable. \eqref{eq:inlam1} is in $\Lambda_1$, and the sum of the remaining terms is in $\Lambda_2$ because
\begin{align*}
	&\quad\qquad\E[ SR_\pre q_\pre(Z,X,1) e_\pre(1) \\
	&\quad\qquad\;\;+ (1-S)\{R_0 q_0(Z,X)e_0 - R_\pre q_\pre(Z,X,0)e_{\pre}(0)\} | Z,X,S,R_0,R_\pre ] = 0, \\
	&\E[ q_\pre(Z,X,S) e_\pre^2(S) | Z,X,S,R_\pre=1 ] \in \overline{R(F_\pre)}, \quad
	\E[ q_0(Z,X)e_0^2 | Z,X,S=0,R_0=1 ] \in \overline{R(F)}.
\end{align*}
The first equality follows by iterated expectation from the equations in Theorem \ref{thm:CPTAL_marl} and Assumptions \ref{A:proxy_cond_marl} and \ref{A:MARL}.

Consistency of the POR and the PIPW estimators under $\mathcal{M}_{\mathcal{H}}$ and $\mathcal{M}_{\mathcal{Q}}$, respectively, is direct from Theorem \ref{thm:CPTAL_marl}. 
Double robustness of $\hat\tau_{\textrm{PDR}}$ is direct from \eqref{eq:hpre}-\eqref{eq:eta}. Now, assume the model $\mathcal M_{\mathcal H} \cap \mathcal M_{\mathcal Q}$ and define $\varphi = \psi + \tau$ so that $\hat\tau^{\textrm{PDR}} = \pr_n\{\hat\varphi\}$. We have $\sqrt n(\hat\tau^{\textrm{PDR}} - \tau) 
   = \sqrt{n}(\pr_n-\pr)\{\varphi\}
     + \sqrt{n}(\pr_n-\pr)\{\hat\varphi - \varphi\}
     + \sqrt{n}\pr\{\hat\varphi - \varphi\}$.
The first term converges in distribution to $\mathcal N(0, \E[\psi^2])$ as $n\to\infty$. The second and the third terms are $o_P(1)$ by Lemma 2 in \cite{kennedy2020sharp} and 
\begin{align*}
	&|\pr\{\hat\varphi - \varphi\}| \\
 	&\le \left[K_1 \{ \|F'(\hat{q}_0-q_0)\| \cdot (\|\hat{h}_0-h_0\| + \|\hat{h}_\pre-h_\pre\|) + \|F'_\pre(\hat q_\pre - q_\pre)\| \|\hat{h}_\pre - h_\pre\| \}\right] \\
 	&\quad \wedge \left[K_2 \{ \|\hat{q}_0-q_0\| \cdot (\|F(\hat{h}_0-h_0)\| + \|F_\pre(\hat{h}_\pre-h_\pre)\|) + \|\hat q_\pre - q_\pre\| \|F_\pre(\hat{h}_\pre-h_\pre)\| \}\right]
\end{align*}
for finite constants $K_1$, $K_2$. Hence, $\sqrt n(\hat\tau^{\textrm{PDR}} - \tau)$ is asymptotically normal with mean $0$ and variance $\E[\psi^2]$ in $\mathcal M_{\mathcal H} \cap \mathcal M_{\mathcal Q}$.
\end{proof}

\section{Supplementary Simulation Details and Results}   \label{sec:sim_detail}
\tred{\subsection{Details of the Data-Generating Process}}  \label{AA:dgp}
For the simulation study, we generated $S, W, Z, Y_\pre$, and $Y_0$ with the following linear models:
\begin{align*}
S\mid \boldsymbol X,U
&\sim\mathrm{Bernoulli}\left[
\operatorname{expit}\left\{
(\boldsymbol X^\top,U)\beta_S+2\mathbf{1}(\gamma=2)
\right\}\right],\\
W&=(\boldsymbol X^\top,U)\beta_W+\varepsilon_W,\\
Z&=(\boldsymbol X^\top,U)\beta_Z
+2S(\boldsymbol X^\top,U)(\beta_S+0.2\boldsymbol{1})
+\varepsilon_Z,\\
Y_{\mathrm{pre}}
&=(\boldsymbol X^\top,U)\beta_Y
+3S(\boldsymbol X^\top,U)(\beta_S-0.25\boldsymbol{1})
+\varepsilon_{\mathrm{pre}},\\
Y_0
&=(\boldsymbol X^\top,U)(\beta_Y+4.5\boldsymbol{1})
+3S(\boldsymbol X^\top,U)(\beta_S-0.25\boldsymbol{1})
+\varepsilon_0,
\end{align*}
where
$\varepsilon_W,\varepsilon_Z \overset{\mathrm{ind}}{\sim}\mathcal N(0,0.75)$ and $\varepsilon_{\mathrm{pre}},\varepsilon_0 \overset{\mathrm{ind}}{\sim}\mathcal N(0,1)$.
The coefficients are generated as 
\[
\begin{aligned}
\beta_Y = 3.75\,\frac{V_Y}{\|V_Y\|_2}, \quad
\beta_W = 5.25\,\frac{V_W}{\|V_W\|_2}, \quad
\beta_Z = 6\,\frac{V_Z}{\|V_Z\|_2}, \quad
\beta_S = 0.5\,\frac{V_S}{\|V_S\|_2},
\end{aligned}
\]
where \(V_Y,V_W,V_Z,V_S\in\mathbb R^3\) are independently generated random vectors with i.i.d. standard uniform entries. The vector \(V_S\) is generated using a fixed random seed of 2. To induce misspecification of the bridge function models, we replaced \(W\) and \(X\) with \((W+3)^{2/3}\) and \((X+3)^{2/3}\), respectively.

\subsection{Computational Resources}  \label{sec:compute_resc}
All experiments were conducted on an Apple M1 Pro machine equipped with 16 GB of RAM. 
The average computational runtime per simulation iteration was approximately 13.5 seconds. With eight distinct scenarios evaluated, the total runtime per scenario averaged around 110 minutes. 
The cumulative computational resources utilized throughout the broader project, including preliminary testing and exploratory analyses, substantially exceeded the runtimes reported for these final evaluations.

\subsection{Results from Section \ref{sec:sim}}
Table \ref{table:simresult1} supplements Figure \ref{fig:simres} by reporting bias, root mean squared error, empirical coverage probability, and the average length of the 95\% confidence interval.
\begin{table*}[h]
\caption{Summary of simulation results. Bias, root mean square error (RMSE), empirical coverage probability (CP), and 95\% confidence interval length (CI len.) are evaluated.}
\
\label{table:simresult1}
\centering
\begin{small}
\begin{sc}
\setlength{\tabcolsep}{4.3pt}
\renewcommand{\arraystretch}{0.9}
\begin{tabular}{rrrrrrrrrr}
\toprule
     &  & \multicolumn{4}{c}{$\gamma=1$}          & \multicolumn{4}{c}{$\gamma=2$}         \\
\cmidrule(lr){3-6} \cmidrule(lr){7-10}
     &      & Bias   & RMSE  & CP    & CI len. & Bias   & RMSE  & CP    & CI len. \\
\midrule
    & PDR  & -0.005 & 0.333 & 0.918 & 0.725   & -0.115 & 0.522 & 0.918 & 1.662   \\
    Correct $d, q$ 
     & PIPW & -0.013 & 0.327 & 0.918 & 0.7    & -0.117 & 0.5 & 0.872 & 1.556  \\
    & POR  & -0.069 & 0.385 & 0.898 & 0.95   & -0.319 & 0.594 & 0.882 & 1.916  \\[0.1cm]
    & PDR  & -0.005 & 0.333 & 0.942 & 0.736   & -0.063 & 0.538 & 0.922 & 1.742  \\
    Incorrect $d$ 
     & PIPW & -0.012 & 0.327 & 0.928 & 0.707   & -0.129 & 0.536 & 0.884 & 1.546  \\
    & POR  & 1.542 & 2.729 & 0.884 & 8.52   & 2.449 & 3.802 & 0.86 & 10.562  \\[0.1cm]
    & PDR  & -0.013 & 0.331 & 0.918 & 0.728   & -0.174 & 0.537 & 0.896 & 1.685  \\
    Incorrect $q$ 
     & PIPW & -0.294 & 0.549 & 0.864 & 1.64   & -0.916 & 1.085 & 0.464 & 1.852  \\
    & POR  & -0.058 & 0.345 & 0.908 & 0.887   & -0.322 & 0.6 & 0.878 & 1.906  \\[0.1cm]
    & PDR  & 0.13 & 0.518 & 0.912 & 1.716   & 0.426 & 1.075 & 0.932 & 3.801  \\
    Incorrect $d, q$ 
     & PIPW & -0.328 & 0.59 & 0.858 & 1.64   & -0.907 & 1.059 & 0.474 & 1.864  \\
    & POR  & 1.56 & 2.663 & 0.864 & 8.502   & 2.237 & 3.496 & 0.882 & 10.481 \\[0.15cm]
    & MB & -0.38 & 0.482 & 0.296 & 0.621   & -1.267 & 1.342 & 0.082 & 1.338   \\
    & SDID & -1.387 & 1.439 & 0. & 0.804   & -1.565 & 1.657 & 0.092 & 1.694    \\
    & MC-NNM & -0.738 & 0.798 & 0.008 & 0.648   & -1.463 & 1.546 & 0. & 0.263    \\
\bottomrule
\end{tabular}
\end{sc}
\end{small}
\end{table*}

\tred{\subsection{Additional Experiment with Missing Outcomes}}
We generated $\boldsymbol{X}$, $U$, $W$, $Z$, $Y_{\pre}$, and $Y_0$ according to a data-generating mechanism similar to that in Section \ref{sec:sim}. The distinctions are as follows:
the untransformed unmeasured confounder is instead generated from $\tilde U\sim\mathcal N(0.75,0.1)$, and $S, Y_\pre, Y_0$ are generated as
\begin{align*}
S \mid \boldsymbol X,U &\sim \mathrm{Bernoulli}\left(
    \operatorname{expit}\left\{(\boldsymbol X^\top,U)\beta_S\right\}
\right),\\
Y_{\mathrm{pre}} &= (\boldsymbol X^\top,U)\beta_Y
    +1.2S(X^\top,U)(\beta_S+0.1\boldsymbol{1})
    +\varepsilon_{\mathrm{pre}},\\
Y_0 &= (\boldsymbol X^\top,U)(\beta_Y+1.5\boldsymbol{1})
    +1.2S(\boldsymbol X^\top,U)(\beta_S+0.1\boldsymbol{1})
    +\varepsilon_0,
\end{align*}
where $\varepsilon_W,\varepsilon_Z \overset{\mathrm{ind}}{\sim}\mathcal N(0,0.5)$ and $\varepsilon_{\mathrm{pre}},\varepsilon_0 \overset{\mathrm{ind}}{\sim}\mathcal N(0,0.25)$.
The missingness indicators $R_{\pre}$ and $R_0$ were generated independently using logistic models depending on $\boldsymbol{X}$ and $U$, each with the same missingness rates across $t=\pre,0$ of approximately $5\%$ ($\rho=1$) or $20\%$ ($\rho=0.2$):
\begin{align*}
R_0 \mid \tilde X_1,\tilde X_2,\tilde U
&\sim \mathrm{Bernoulli}\left(
    \operatorname{expit}\left\{
    1+\rho(2\tilde X_1+2\tilde X_2+0.5\tilde U)
    \right\}
\right),\\
R_{\mathrm{pre}} \mid \tilde X_1,\tilde X_2,\tilde U
&\sim \mathrm{Bernoulli}\left(
    \operatorname{expit}\left\{
    1+\rho(2\tilde X_1+2\tilde X_2+0.5\tilde U)
    \right\}
\right),
\end{align*}
The coefficients are generated as
\begin{align*}
	\beta_Y = 1.25\,\frac{V_Y}{\|V_Y\|_2}, \quad
	\beta_W = 1.75\,\frac{V_W}{\|V_W\|_2}, \quad
	\beta_Z = 2\,\frac{V_Z}{\|V_Z\|_2}, \quad
	\beta_S = 0.5\,\frac{V_S}{\|V_S\|_2}.
\end{align*}

The bridge functions \eqref{eq:hpre}-\eqref{eq:eta} were estimated using neural networks with the untransformed covariates $\tilde{\boldsymbol{X}}$, following an optimization procedure similar to that described in Appendix \ref{sec:learning}. 
Although sample splitting is required for the theoretical guarantees, we used the same sample for nuisance function estimation and computation of the empirical means. 
To induce misspecification of the bridge function models, we replaced \(W\) and \(X\) with \((W+3)^{2/3}\) and \((X+3)^{2/3}\), respectively.

We evaluated PDR \eqref{eq:PDR_marl}, PIPW \eqref{eq:PIPW_marl}, and POR \eqref{eq:POR_marl}, considering all combinations of correct and incorrect specification of $\mathcal{H}$ and $\mathcal{Q}$. For each scenario, we conducted $500$ simulation replications with a sample size of $n=500$. Tables \ref{table:simresult1_marl} and \ref{table:simresult2_marl} summarize the results. 
\tred{We note that the outcomes are generated on different scales from those in Section \ref{sec:sim}, precluding direct comparison of the results.}

\clearpage

\begin{table}[hbt!]
\caption{Simulation result with missing outcomes ($\gamma=1$). Bias, root mean square error (RMSE), empirical coverage probability (CP), and 95\% confidence interval length (CI len.) are evaluated under varying missingness rates ($\pr(R_t=0)$).}
\label{table:simresult1_marl}
\centering
\begin{small}
\begin{sc}
\setlength{\tabcolsep}{4.3pt}
\renewcommand{\arraystretch}{0.9}
\begin{tabular}{rrrrrrrrrr}
\toprule
$\pr(R_t=0)$                 &      & \multicolumn{4}{c}{5\%}          & \multicolumn{4}{c}{20\%}         \\
\cmidrule(lr){3-6} \cmidrule(lr){7-10}
                             &      & Bias   & RMSE  & CP    & CI len. & Bias   & RMSE  & CP    & CI len. \\
\midrule
\multirow{3}{*}{Correct $\mc{H},\mc{Q}$} & PDR  & 0.027  & 0.122 & 0.932 & 0.475   & 0.026  & 0.144 & 0.914 & 0.552   \\
                             & PIPW & -0.113  & 0.302 & 0.916 & 1.077   & -0.053  & 0.383 & 0.950 & 1.505   \\
                             & POR  &  0.033 & 0.491 & 0.934 & 1.792   & 0.007 & 0.517 & 0.924 & 1.831   \\[0.1cm]
\multirow{3}{*}{Incorrect $\mc{H}$}   & PDR  & -0.053 & 0.215 & 0.914 & 0.769   & -0.002 & 0.263 & 0.952 & 1.034   \\
                             & PIPW & -0.097 & 0.308 & 0.920 & 1.091   & -0.032 & 0.382 & 0.950 & 1.512   \\
                             & POR  & -2.458 & 2.472 & 0. & 1.031   & -2.453 & 2.468 & 0. & 1.046   \\[0.1cm]
\multirow{3}{*}{Incorrect $\mc{Q}$}   & PDR  & 0.066  & 0.190 & 0.940 & 0.724   & 0.077  & 0.239 & 0.908 & 0.822   \\
                             & PIPW & -0.957 & 1.013 & 0.214 & 1.360   & -1.178  & 1.226 & 0.082 & 1.325   \\
                             & POR  & 0.063  & 0.462 & 0.956 & 1.786   & 0.042 & 0.512 & 0.930 & 1.789   \\[0.1cm]
\multirow{3}{*}{Incorrect $\mc{H},\mc{Q}$}   & PDR  & -0.572 & 0.616 & 0.302 & 0.887   & -0.688 & 0.725 & 0.150 & 0.888   \\
                             & PIPW & -0.947 & 1.015 & 0.250 & 1.375   & -1.155 & 1.211 & 0.104 & 1.317   \\
                             & POR  & -2.448 & 2.463 & 0. & 1.039   & -2.434 & 2.450 & 0. & 1.033  \\
\bottomrule
\end{tabular}
\end{sc}
\end{small}
\end{table}

\begin{table}[]
\caption{Simulation result with missing outcomes ($\gamma=2$). Bias, root mean square error (RMSE), empirical coverage probability (CP), and 95\% confidence interval length (CI len.) are evaluated under varying missingness rates ($\pr(R_t=0)$).}
\label{table:simresult2_marl}
\centering
\begin{small}
\begin{sc}
\setlength{\tabcolsep}{4.3pt}
\renewcommand{\arraystretch}{0.9}
\begin{tabular}{rrrrrrrrrr}
\toprule
$\pr(R_t=0)$                 &      & \multicolumn{4}{c}{5\%}          & \multicolumn{4}{c}{20\%}         \\
\cmidrule(lr){3-6} \cmidrule(lr){7-10}
                             &      & Bias   & RMSE  & CP    & CI len. & Bias   & RMSE  & CP    & CI len. \\ 
\midrule
\multirow{3}{*}{Correct $\mc{H},\mc{Q}$} & PDR  & 0.017  & 0.121 & 0.948 & 0.483   & 0.051  & 0.145 & 0.930 & 0.541   \\ 
                             & PIPW & -0.097 & 0.233 & 0.910 & 0.852   & -0.034 & 0.271 & 0.952 & 1.159   \\ 
                             & POR  & -0.189 & 0.408 & 0.882 & 1.332   & -0.213 & 0.426 & 0.880 & 1.329   \\[0.1cm] 
\multirow{3}{*}{Incorrect $\mc{H}$}   & PDR  & -0.059 & 0.182 & 0.920 & 0.676   & -0.011 & 0.247 & 0.920 & 0.874   \\
                             & PIPW & -0.107 & 0.241 & 0.896 & 0.846   & -0.051 & 0.325 & 0.918 & 1.134   \\
                             & POR  & -2.062 & 2.070 & 0. & 0.727   & -2.056 & 2.067 & 0. & 0.722   \\[0.1cm]
\multirow{3}{*}{Incorrect $\mc{Q}$}   & PDR  & -0.037 & 0.165 & 0.936 & 0.627   & -0.015 & 0.186 & 0.922 & 0.708   \\
                             & PIPW & -0.788 & 0.835 & 0.178 & 1.051   & -0.950 & 0.993 & 0.104 & 1.034   \\
                             & POR  & -0.207 & 0.401 & 0.880 & 1.323   & -0.192 & 0.409 & 0.880 & 1.349   \\[0.1cm]
\multirow{3}{*}{Incorrect $\mc{H},\mc{Q}$}   & PDR  & -0.526 & 0.563 & 0.244 & 0.770   & -0.629 & 0.661 & 0.160 & 0.782   \\
                             & PIPW & -0.795 & 0.841 & 0.180 & 1.061   & -0.955 & 0.994 & 0.098 & 1.051   \\
                             & POR  & -2.043 & 2.051 & 0. & 0.724   & -2.065 & 2.073 & 0. & 0.723 \\
\bottomrule 
\end{tabular}
\end{sc}
\end{small}
\end{table}

\clearpage


\newpage
\section*{NeurIPS Paper Checklist}

\begin{enumerate}

\item {\bf Claims}
    \item[] Question: Do the main claims made in the abstract and introduction accurately reflect the paper's contributions and scope?
    \item[] Answer: \answerYes{} 
    \item[] Justification: Our main claims accurately reflect both the theoretical framework presented in Sections 2-4 and the experimental results detailed in Section 5.

\item {\bf Limitations}
    \item[] Question: Does the paper discuss the limitations of the work performed by the authors?
    \item[] Answer: \answerYes{} 
    \item[] Justification: Assumptions in Sections 2-4, while outlining standard conditions in the field, discuss scenarios that may not be suitable for the proposed method. 
    	Appendix \ref{sec:cptl_test}  clarifies the limitations of the proposed falsification test.

\item {\bf Theory assumptions and proofs}
    \item[] Question: For each theoretical result, does the paper provide the full set of assumptions and a complete (and correct) proof?
    \item[] Answer: \answerYes{} 
    \item[] Justification: We provided the necessary assumptions in Sections 2-4, and the regularity conditions along with complete proofs in the Appendix. 

    \item {\bf Experimental result reproducibility}
    \item[] Question: Does the paper fully disclose all the information needed to reproduce the main experimental results of the paper to the extent that it affects the main claims and/or conclusions of the paper (regardless of whether the code and data are provided or not)?
    \item[] Answer: \answerYes{} 
    \item[] Justification: Settings to reproduce the results are provided in Section 5 and the Appendix. 
    	We have made the code used to generate our results available.

\item {\bf Open access to data and code}
    \item[] Question: Does the paper provide open access to the data and code, with sufficient instructions to faithfully reproduce the main experimental results, as described in supplemental material?
    \item[] Answer: \answerYes{} 
    \item[] Justification: We have made the code used to generate our results available.

\item {\bf Experimental setting/details}
    \item[] Question: Does the paper specify all the training and test details (e.g., data splits, hyperparameters, how they were chosen, type of optimizer) necessary to understand the results?
    \item[] Answer: \answerYes{} 
    \item[] Justification:  We provided the experimental scheme in Section \ref{sec:sim} and Appendix \ref{sec:sim_detail}.

\item {\bf Experiment statistical significance}
    \item[] Question: Does the paper report error bars suitably and correctly defined or other appropriate information about the statistical significance of the experiments?
    \item[] Answer: \answerYes{} 
    \item[] Justification: Box plots from the experiment are reported, with whiskers extending to the farthest data point within 1.5 times IQR from the box. Standard errors as well as coverage rate are reported in Table \ref{table:simresult1}. For semi-synthetic experiment, bias and standard error are reported in Table \ref{table:lzanresult}.

\item {\bf Experiments compute resources}
    \item[] Question: For each experiment, does the paper provide sufficient information on the computer resources (type of compute workers, memory, time of execution) needed to reproduce the experiments?
    \item[] Answer: \answerYes{} 
    \item[] Justification: Appendix \ref{sec:compute_resc} details the computational resources used and the average execution time for each task in the simulation study.
    
\item {\bf Code of ethics}
    \item[] Question: Does the research conducted in the paper conform, in every respect, with the NeurIPS Code of Ethics \url{https://neurips.cc/public/EthicsGuidelines}?
    \item[] Answer: \answerYes{} 
    \item[] Justification: The research conducted in the paper conforms with the code of ethics in every aspect.

\item {\bf Broader impacts}
    \item[] Question: Does the paper discuss both potential positive societal impacts and negative societal impacts of the work performed?
    \item[] Answer: \answerYes{} 
    \item[] Justification: Sections 2 and 3 discuss the potential positive societal impact.The proposed methodology carries minimal risk of negative societal impacts, provided it is not misapplied to data-generating processes that fundamentally violate the identifying assumptions outlined in Sections 2–4. To address potential concerns regarding practical robustness, we provide an extensive sensitivity analysis demonstrating the estimator's reliability under controlled assumption violations.
    
\item {\bf Safeguards}
    \item[] Question: Does the paper describe safeguards that have been put in place for responsible release of data or models that have a high risk for misuse (e.g., pre-trained language models, image generators, or scraped datasets)?
    \item[] Answer: \answerNA{} 
    \item[] Justification: The methodology presented in the paper does not pose high risk for misuse. To address potential concerns regarding practical robustness, we provide an extensive sensitivity analysis demonstrating the estimator's reliability under controlled assumption violations.

\item {\bf Licenses for existing assets}
    \item[] Question: Are the creators or original owners of assets (e.g., code, data, models), used in the paper, properly credited and are the license and terms of use explicitly mentioned and properly respected?
    \item[] Answer: \answerYes{} 
    \item[] Justification: The data used, and its original publication are explicitly cited in Section 5.

\item {\bf New assets}
    \item[] Question: Are new assets introduced in the paper well documented and is the documentation provided alongside the assets?
    \item[] Answer: \answerYes{} 
    \item[] Justification: The code used to reproduce the results presented in this study has been provided as structured Jupyter notebooks.

\item {\bf Crowdsourcing and research with human subjects}
    \item[] Question: For crowdsourcing experiments and research with human subjects, does the paper include the full text of instructions given to participants and screenshots, if applicable, as well as details about compensation (if any)? 
    \item[] Answer: \answerNA{} 
    \item[] Justification: This paper does not involve crowdsourcing nor research with human subjects.

\item {\bf Institutional review board (IRB) approvals or equivalent for research with human subjects}
    \item[] Question: Does the paper describe potential risks incurred by study participants, whether such risks were disclosed to the subjects, and whether Institutional Review Board (IRB) approvals (or an equivalent approval/review based on the requirements of your country or institution) were obtained?
    \item[] Answer: \answerNA{} 
    \item[] Justification: This paper does not involve crowdsourcing nor research with human subjects.

\item {\bf Declaration of LLM usage}
    \item[] Question: Does the paper describe the usage of LLMs if it is an important, original, or non-standard component of the core methods in this research? Note that if the LLM is used only for writing, editing, or formatting purposes and does \emph{not} impact the core methodology, scientific rigor, or originality of the research, declaration is not required.
    \item[] Answer: \answerNA{} 
    \item[] Justification: This research does not involve the usage of LLMs in any important, original, or non-standard components of the core methods.

\end{enumerate}

\end{document}